# Quantitative bias analysis for outcome phenotype error correction in comparative effect estimation: an empirical and synthetic evaluation

James Weaver, Patrick B. Ryan, Victoria Y. Strauss, Marc A. Suchard, Joel Swerdel, Daniel Prieto-Alhambra

Centre for Statistics in Medicine, Nuffield Department of Orthopedics, Rheumatology and Musculoskeletal Sciences, University of Oxford, Oxford, UK (James Weaver, Daniel Prieto-Alhambra); Observational Health Data Analytics, Janssen Research and Development, Titusville, NJ, USA (James Weaver, Patrick B. Ryan, Joel Swerdel); Observational Health Data Sciences and Informatics, New York, NY, USA (James Weaver, Patrick B. Ryan, Victoria Y. Strauss, Joel Swerdel, Marc A. Suchard); Department of Biomedical Informatics, Columbia University Medical Center, New York, NY, USA (Patrick B. Ryan); Global Biostatistics & Data Sciences, Boehringer Ingelheim Pharma GmbH & Co. KG., Ingelheim, Germany (Victoria Y. Strauss); Department of Biostatistics, Fielding School of Public Health, and Department of Biomathematics, David Geffen School of Medicine, UCLA, Los Angeles, CA, USA (Marc A. Suchard); Medical Informatics, Erasmus Medical Centre, Rotterdam, Netherlands (Daniel Prieto-Alhambra)

This work was partially supported by the Clarendon Fund and Brasenose College Scholarship, University of Oxford (JW). This work was partially funded by the UK National Institute for Health Research (DPA).

JW and DPA led the conception and design of the work, and all authors made substantial contributions; PBR led the acquisition of the data; JW led the data analysis, and all authors made substantial contributions. All authors were substantially involved in results interpretation; all authors have contributed to the drafting and critical revision of the manuscript for important intellectual content; all authors have given final approval and agree to be accountable for all aspects of the work.

The data that support the findings of this study are available from IBM® Marketscan® and Optum® but restrictions apply to the availability of these data, which were used under license and so are not publicly available. Aggregated (i.e., no person-level data) results that are the basis of the study findings are publicly available at https://github.com/OHDSI/ShinyDeploy/tree/master/QbaEvaluation/data.

**JW**, **PR**, and **JS** are employees of shareholders of Janssen R&D (a Johnson & Johnson company). **MAS** receives grant support from the US National Institutes of Health, US Food & Drug Administration and US Department of Veterans Affairs and contracts from Janssen R&D. **VS** is an employee of Boehringer Ingelheim Pharma GmbH. **DPA**'s department has received grant/s from Amgen, Chiesi-Taylor, Lilly, Janssen, Novartis, and UCB Biopharma. His research group has received consultancy fees from Astra Zeneca and UCB Biopharma. Amgen, Astellas, Janssen, Synapse Management Partners and UCB Biopharma have funded or supported training programs organized by **DPA**'s department.

We thank Dr. Joshua Gagne for his thoughtful advice.

## Abstract

Outcome phenotype measurement error is rarely corrected in comparative effect estimation studies in observational pharmacoepidemiology. Quantitative bias analysis (QBA) is a misclassification correction method that algebraically adjusts person counts in exposure-outcome contingency tables to reflect the magnitude of misclassification. The extent QBA minimizes bias is unclear because few systematic evaluations have been reported. We empirically evaluated QBA impact on odds ratios (OR) in several comparative effect estimation scenarios. We estimated non-differential and differential phenotype errors with internal validation studies using a probabilistic reference. Further, we synthesized an analytic space defined by outcome incidence, uncorrected ORs, and phenotype errors to identify which combinations produce invalid results indicative of input errors. We evaluated impact with relative bias [(OR-$OR_{QBA}$)]/OR×100%]. Results were considered invalid if any contingency table cell was corrected to a negative number. Empirical bias correction was greatest in lower incidence scenarios where uncorrected ORs were larger. Similarly, synthetic bias correction was greater in lower incidence settings with larger uncorrected estimates. The invalid proportion of synthetic scenarios increased as uncorrected estimates increased. Results were invalid in common, low incidence scenarios indicating problematic inputs. This demonstrates the importance of accurately and precisely estimating phenotype errors before implementing QBA in comparative effect estimation studies.

## Abbreviations

| | |
|---|---|
| ACEI | angiotensin-converting enzyme |
| ARB | angiotensin receptor blockers |
| CDM | common data model |
| CI | confidence interval |
| COVID-19 | coronavirus disease 2019 |
| DOD | date of death |
| EHR | electronic health record |
| HR | hazard ratio |
| MDCD | Medicaid |
| MDCR | Medicare |
| NPV | negative predictive value |
| OHDSI | Observational Health Data Sciences and Informatics |
| OMOP | Observational Medical Outcomes Partnership |
| OR | odds ratio |
| PPV | positive predictive value |
| PS | propensity score |
| QBA | quantitative bias analysis |
| SE | standard error |
| TAR | time-at-risk |
| THZ | thiazide/thiazide-like diuretics |

Outcome phenotype measurement error is acknowledged but rarely corrected in observational comparative effect estimation[1, 2]. Literature is available on methods to correct observed confounding from selective treatment assignment using propensity scores (PS), much of which has been studied, endorsed, and routinely used in research using large healthcare databases[3-7]. However, despite availability and endorsement, quantitative assessment of the direction, magnitude, and uncertainty from information bias is rarely included clinical pharmacoepidemiology studies[1, 2].

Phenotyping accuracy in healthcare databases depends on the clinical and behavioral attributes of the underlying clinical concept and the content completeness and reliability in the database from which cases are identified. Good phenotype development requires understanding patient care-seeking behavior, diagnosis and treatment patterns, and construction and characteristic details about the database[8]. Appeals have been made for researchers to use quantitative bias analysis (QBA) to identify and incorporate the phenotype algorithm measurement error in their studies[9].

Phenotype errors occur when data required for a definition are missing, incorrectly recorded, or when discordance exists between a physiological description and its operationalization in a database for research. An example is ascertaining coronavirus disease 2019 (COVID-19) cases in administrative claims databases with linkages to laboratory testing information. Often patients with a COVID-19 diagnosis also have varying patterns of positive and negative Covid-19 testing results temporally proximal to their diagnosis date, which complicates the process of accurately defining cases. Some patients have an observed diagnosis code without a confirmatory laboratory result, either because a test was not performed, or the result was not recorded. Alternatively, some patients have a positive test observed without an associated diagnosis. Outcome phenotype definition errors can lead to information bias. Moreover, in distributed database network studies a single outcome phenotype definition is generally applied across multiple, disparate databases that may vary in terms of purpose, data capture process, population and temporal coverage, and clinical content. Given database heterogeneity, outcome phenotype errors will also vary.

QBA includes a method for measuring and correcting information bias from outcome phenotype error[10]. It is a deterministic function that takes the observed contingency table of exposure-outcome status and misclassification parameters (either (sensitivity and specificity) or (positive predictive value[PPV] and negative predictive value[NPV])) as input to adjust the observed effect estimate and produce a 'corrected' effect estimate intended to better reflect the true underlying parameter. However, inaccuracies or imprecision in the inputs (such as an improper estimate of specificity) can result in the QBA algebra yielding implausible values, to the extreme of patient counts being corrected to negative numbers.

A review covering 2006-2019[1] found that QBA use is rare but increasing, methodological reporting was often incomplete, and bias correction varied across studies and methods. A probabilistic QBA review in pharmacoepidemiology[2] reported that use is rare and methodological approaches and reporting are suboptimal. Given underutilization and implementation difficulties[9, 11], QBA operating performance is unclear. This is the basis for our QBA evaluation. The extent to which QBA procedures can be readily applied in the context of pharmacovigilance studies involving low incidence events, given the challenges of producing accurate and precise measurement error estimates, has not been thoroughly explored.

## Objective

Our intent was twofold. First, we empirically evaluated QBA for outcome phenotype error correction in several illustrative comparative effect estimation scenarios. Second, we synthesized an analytic space defined by fixed outcome incidence proportions (hereafter incidence), uncorrected effect estimates, and outcome phenotype measurement errors to identify the range for which inputs would be reasonable for QBA application.

# METHODS

The pre-specified protocol is available at https://ohdsi-studies.github.io/QbaEvaluation/protocol.html (EUPAS104459) and the study source code is available at https://github.com/ohdsi-studies/QbaEvaluation.

## Quantitative bias analysis

Simple QBA for outcome misclassification algebraically rearranges the observed distribution of a categorical contingency table of exposure-outcome patient counts to reflect the true distribution given no outcome phenotype error. Equations that relate effect estimate calculation to the outcome phenotype errors are used and standard errors (SE) are adjusted[12, 13] (**Web Tables 1 and 2**). Multidimensional QBA is an extension of simple QBA where phenotyping errors are uncertain, so a range of plausible errors are used as input to produce a corresponding range of QBA-corrected estimates. A popular heuristic states that outcome error non-differential to exposure status will bias effect estimates towards the null, although this assumption does not hold in all scenarios[14-16]. Provided exposure-differential phenotype errors are attainable, QBA methods can incorporate them. We implemented simple QBA using sensitivity and specificity inputs for illustration. A limitation of QBA is that certain combinations of incidence, uncorrected effect estimate, and measurement errors can produce a corrected contingency table with negative cell counts, from which a corrected effect estimate cannot be computed[10]. We refer to such inputs as invalid. Certain input scenarios can result in a corrected contingency table where all cell counts are non-negative, but the incidence becomes such that the corrected estimate and SE becomes large.

## Probabilistic reference standard validation

Validating phenotypes follows a process: select health condition, choose reference standard against which to assess the phenotype algorithm, develop the algorithm using temporal logic and clinical codes from the database, apply phenotype algorithm to select patients for validation, and assess phenotype algorithm's performance against reference standard[17].

Recent methods development has introduced probabilistic reference standard validation, referred to hereafter as PheValuator[18, 19]. PheValuator is a method to calculate the performance characteristics of phenotype algorithms: sensitivity, specificity, PPV, and NPV. Briefly, PheValuator develops a diagnostic predictive model to assign a probabilistic reference set to compare against a phenotype algorithm for performance assessment. Concisely, it enables rapid internal phenotype validation. More details provided in **Web Appendix 1**.

## Empirical example

We replicated two comparative analyses from a set of new user, active-comparator, cohort studies[20] executed against four US administrative claims and one electronic health record databases (**Web Table 3**) standardized to the Observational Health Data Sciences and Informatics (OHDSI) collaborative's Observational Medical Outcomes Partnership (OMOP) common data model (CDM)[21, 22].

We assessed the risk of ischemic stroke for new-users of angiotensin-converting enzyme inhibitors (ACEI) vs angiotensin receptor blockers (ARB) and vs thiazide/thiazide-like diuretics (THZ) in patients with hypertension during 365- and 730-day time-at-risk (TAR) periods. Detailed exposure and outcome cohort definitions are available in the **Protocol Appendix Section B**. The study we replicated[20] reported no evidence of increased risk for ACEI vs ARB new users (pooled hazard ratio (HR) 1.07, 95% confidence interval [CI] 0.92-1.27) and evidence of increased risk for ACEI vs THZ new users (pooled HR 1.18, 95% CI 1.02-1.40). We chose these comparisons as a negative and positive control, respectively.

For each comparison during each TAR, we estimated the risk for ischemic stroke using six analyses:

1) Unadjusted
2) Non-differential QBA
3) Differential QBA
4) 1:1 PS-matched
5) 1:1 PS-matched, non-differential QBA
6) 1:1 PS-matched, differential QBA

We used large-scale PS to ensure baseline covariate balance on directly and indirectly measured covariates[7, 23]. For QBA, we used exposure-agnostic and exposed population outcome sensitivity and specificity estimates for non-differential and differential analyses, respectively. We executed [5 databases × 2 comparisons × 2 TARs × 6 analyses]=120 analyses. Each analysis produced an exposure-outcome 2x2 table that we used as input for computing odds ratio (OR) estimates and associated 95% CIs.

We assessed QBA performance by calculating bias difference, relative bias, and relative precision between the following 6 analysis comparisons during each TAR:

1) Unadjusted vs non-differential QBA
2) Unadjusted vs differential QBA
3) Non-differential QBA vs. differential QBA
4) PS matched vs PS-matched with non-differential QBA
5) PS matched vs PS-matched with differential QBA
6) PS matched with non-differential QBA vs PS-matched with differential QBA

Bias difference is the difference between the log of the QBA-corrected estimate and the log of the uncorrected estimate [$\log(OR)-\log(OR_{QBA})$]. Relative bias is the percentage change between the QBA-corrected estimate and the uncorrected estimate [$(OR-OR_{QBA})]/OR\times100$]. Relative precision is percentage change in the SE of the estimate, which is related to the width of the CI, after QBA correction [$1/(SE(\log(OR))^2-1/(SE(\log(OR_{QBA}))^2)/(1/(SE(\log(OR))^2\times100$]. For context, if QBA does not change the original estimate, all evaluation metrics equal 0. If QBA changes the original estimate from OR=1 to $OR_{QBA}$=2, then bias difference is log(1)-log(2)=-0.69. We interpret this as the uncorrected OR is half [exp(-0.69)=0.5] of the QBA-corrected OR, or that phenotype error reduced the “true” OR by one half. Further, the relative bias is (1-2)/1*100=-100% which is interpreted as QBA correction reducing the original estimate by 100%. If QBA changes the original estimate SE from SE(log(OR))=0.05 to $SE(\log(OR_{QBA}))$=0.2, then the relative precision is $(1/0.05^2-1/0.2^2)/(1/0.05^2)\times100$=93.8%. We interpret this as QBA correction decreased precision by 94%. These metrics are also applied when comparing non-differential to differential QBA analyses.

We also applied non-differential multidimensional QBA across a range of sensitivity and specificity values (0-1 by $10^{-4}$) to identify phenotype error thresholds where QBA produces negative cell counts.

## Synthetic grid space

We synthesized a grid space of all combinations of four input parameters that cover: 5 outcome incidences [$10^{-1}$, $10^{-2}$, $10^{-3}$, $10^{-4}$, $10^{-5}$], 6 uncorrected ORs [1, 1.25, 1.50, 2, 4, 10], 20 outcome sensitivity values [0.05 to 1.00 by 0.05], and 20 outcome specificity values. Because specificity precision depends on incidence, we generated specificity values within each incidence level. That is, we defined specificity values as 1-incidence to 1.00 by 20 equal increments. For example, where incidence=0.1, the specificity values range from 0.9 to 1 by 0.0526. The complete grid space consisted of 12,000 2x2 tables, each with 1,000,000 target and 1,000,000 comparator exposures. We chose large, equally sized exposure groups to focus assessment on the direction and magnitude of bias correction. For each incidence-OR stratum, we computed a distribution of QBA-corrected ORs across combinations of sensitivity and specificity values and plotted their contours across the complete incidence-OR grid space. We reported QBA-corrected OR point estimates and associated sensitivity and specificity at the minimum (uncorrected OR where sensitivity=specificity=1), 25$^{th}$, 50$^{th}$, and 75$^{th}$ percentiles, and maximum of the distribution.

We assessed QBA performance by calculating bias difference and relative bias between the uncorrected OR and 25$^{th}$, 50$^{th}$, and 75$^{th}$ percentile and maximum QBA-corrected OR point estimates stratified by incidence. We interpret analyses that produce negative corrected cell counts as having invalid QBA input combinations where the uncorrected OR would be unobservable. Alternatively, there is no true OR given such incidence, sensitivity, and specificity values that would produce the uncorrected OR.

# RESULTS

All empirical and synthetic results are available in a web-based, interactive **Web Application** at https://data.ohdsi.org/QbaEvaluation/.

## Empirical example

Results from the empirical example are available in the Simple QBA tab of the **Web Application**. User instructions are in **Web Appendix 2**.

Across databases, sensitivity was greater among exposed populations than exposure-agnostic populations. In exposure-agnostic populations, sensitivity ranged from 0.3797 (Optum® de-identified Electronic Health Record Database[Optum EHR®]) to 0.6209 (IBM MDCR) and specificity ranged from 0.9967 (Optum's de-identified Clinformatics® Data Mart Database[Clinformatics®]) to 0.9994 (Optum EHR®). Among exposed populations, sensitivity ranged from 0.4783 (ARB in Optum EHR®) to 0.6742 (THZ in IBM MDCR) and specificity ranged from 0.9872 (THZ in IBM MDCR) to 0.9987 (ACEI in Optum EHR®). **Table 1** reports exposure-agnostic and exposed population phenotype errors.

**Figure 1** shows that applying non-differential QBA to unadjusted and PS-matched analyses increases the odds ratio (OR) and 95% CI width for both comparisons in most databases. Differential QBA correction increases the OR and 95% CI width to a greater extent. **Web Figure 1** shows the 365-day analyses result in fewer observed events during the shorter TAR which leads to fewer valid results with larger ORs and CI widths than the higher-powered 730-day findings.

**Table 2** reports evaluation metrics computed for each exposure comparison in each database individually and averaged across databases. Among aggregated results, relative bias and bias difference was greater in the ACEI vs THZ positive control scenario than in the ACI vs ARB negative control scenario. The greatest bias reduction was for ACEI vs THZ that compared the unadjusted OR vs differential QBA correction (relative bias=-107%, bias difference=-0.68) whereas reduction for ACEI vs ARB was of lesser magnitude (relative bias=-18%, bias difference=-0.12). Similarly, ACEI vs THZ showed the largest precision increase (relative precision=91%).

QBA demonstrated the least bias correction where we compared the PS-matched analysis to the PS-matched with non-differential QBA analyses (ACEI vs THZ: relative bias=-0.6%, bias difference=-0.004; ACEI vs ARB: relative bias=-1.7%, bias difference=-0.02). The lower-power, 365-day TAR evaluation metrics showed directional consistency with the 730-day results although bias correction was greater, yet relative precision change was similar. Bias correction rank ordering among analysis comparisons was like the 730-day results. **Web Table 4** displays the 365-day analysis results that show fewer events during the shorter TAR which leads to fewer valid results with larger ORs and less precision than the higher powered 730-day findings.

**Web Appendix 3** reports non-differential multidimensional QBA results with supporting discussion on high specificity precision requirements in low prevalence settings.

## Synthetic grid space

Results from the synthetic grid space are available in the Synthetic grid space tab of the **Web Application**.

**Figure 2** displays the QBA-corrected OR contours across the grid space. Each panel represents an incidence-OR stratum, where rows are the 5 incidence values and columns are the 6 uncorrected ORs. Within each stratum are 400 combinations of sensitivity and specificity values. These incidence, uncorrected OR, sensitivity, and specificity values are the input for QBA analyses that produced 400 ORs corrected for phenotype error. Corrected ORs are produced for sensitivity-specificity combinations where the cells in the corrected 2x2 table are all non-negative. The minimum and maximum valid point estimates are reported, as are the sensitivity-specificity contours of the 25$^{th}$, 50$^{th}$, and 75$^{th}$ percentiles of the corrected OR distribution.

To illustrate, the upper left panel where incidence is $10^{-1}$ and the uncorrected OR is 1.001 displays the minimum valid OR as the uncorrected OR of 1.001 where sensitivity=specificity=1 in the upper right. At the other extreme, the maximum corrected OR of 1.019 where sensitivity is 0.15 and specificity is 0.9053 is displayed in the lower left. The sensitivity-specificity value contours where the corrected OR is the 25$^{th}$ (corrected OR=1.001), 50$^{th}$ (corrected OR=1.002), and 75$^{th}$ (corrected OR=1.004) percentile of the corrected OR distribution are the blue curves. At this incidence where the uncorrected OR is 1.001, QBA inputs are valid where sensitivity ranges from 0.15 to 1 and specificity ranges from 0.9053 to 1. At this incidence-uncorrected OR input, applying QBA at sensitivity-specificity value combinations outside of this range produces corrected 2x2 tables with negative cell counts and an OR cannot be calculated.

Across the space, the minimum required valid specificity was 0.91 where incidence was $10^{-1}$. Where incidence was $10^{-5}$, the minimum required specificity was 0.9999. At higher incidences, variation among lower sensitivity values demonstrated impact on bias correction, but where incidence was less than $10^{-3}$ only specificity affected correction. At higher uncorrected ORs, valid sensitivity-specificity combinations that produced all non-negative corrected 2x2 tables resulted in implausibly inflated ORs. For example, where incidence is $10^{-1}$ and the uncorrected OR is 4, the maximum valid inputs were sensitivity of 0.20 and specificity of 0.9579 producing a corrected estimate of 198.8.

**Figure 3** displays the estimable proportion (y-axis) of the sensitivity by specificity space across uncorrected OR values (x-axis) stratified by incidence (panels). Estimable proportion is the fraction of 400 input sensitivity and sensitivity values within an incidence-uncorrected OR stratum where QBA-correction produces 2x2 tables where all cells are non-negative.

The largest estimable proportion was 0.95 at incidences of $10^{-2}$, $10^{-3}$, $10^{-4}$ where the uncorrected OR was 1.001. The lowest estimable proportion was 0.165 in the stratum where incidence was $10^{-5}$ and the uncorrected OR was 10. For all input incidence values, there was a strong decreasing trend in estimable proportion as the uncorrected OR was increased. For examples, where incidence was $10^{-1}$ the estimable proportion was 0.86 at an uncorrected OR of 1.001 and 0.21 at an uncorrected OR of 10.

**Table 3** reports bias difference and relative bias between the uncorrected OR and the 50$^{th}$ percentile corrected OR point estimate for each incidence-uncorrected OR stratum. The estimable proportion of is reported, as is the sensitivity and specificity associated with the 50$^{th}$ percentile corrected OR point estimate. At incidence of 0.1, relative bias correction ranged from -0.001% where the uncorrected OR was 1.001 (corrected OR=1.002) to -148.6% where the uncorrected OR was 10 (corrected OR=24.86). At this incidence, sensitivity decreased from 0.6 to 0.3 and specificity increased from 0.95 to 0.99 across increasing uncorrected ORs. At incidence of $10^{-5}$, relative bias correction ranged from 0.1% where the uncorrected OR was 1.001 to -48.9% where the uncorrected OR was 4 (corrected OR=5.96). At this incidence, sensitivity decreased from 0.40 to 0.52 and specificity was consistent at >0.99 across increasing uncorrected ORs. See **Web Appendix 4** for the remaining results.

# DISCUSSION

Our empirical analysis demonstrated that QBA yielded greater bias correction in lower incidence scenarios where uncorrected estimates were larger. Our synthetic analysis demonstrated that bias correction was greater in lower incidence scenarios with larger uncorrected estimates. Our findings suggest QBA may be infeasible in common, low incidence pharmacoepidemiology scenarios.

## Empirical example

This is the first study to use exposure-agnostic and exposure-differential phenotype errors estimated by probabilistic reference validation to correct effect estimates subject to outcome phenotype error in empirical scenarios. We found greater bias correction in higher incidence scenarios where uncorrected estimates were larger.

A well-conducted estimation study that used QBA for outcome phenotype error correction assessed newborn cardiac defects following first trimester gestational antidepressant exposure[24]. Among exposed women in the Medicaid Analytic eXtract database, the PS-stratified relative risk of cardiac defects was 1.06 (95% CI, 0.93-1.22) where probabilistic QBA increased the estimate 4.3% from 1.02 (95% CI 0.90-1.15). Chart adjudication-based internal validation determined an outcome PPV of 0.751[25]. Worth noting is that from 660 requested records from 380 hospitals, 158 (24%) were fulfilled. Given cardiac malformation incidence of <1% in both exposure groups (exposed 460/49527×100=0.93%, unexposed 1496/180422×100=0.83%) the authors calculated distributions of sensitivity (Min=0.5,

Mode=0.75, Max=1.0) and specificity (minimum=0.9985, mode=0.99875, maximum=0.999) corresponding to the non-differential PPV.

Comparable scenarios from our empirical example were ACEI vs ARB and ACEI vs THZ in IBM MDCD under the 730d TAR, PS-matched analysis where we applied non-differential correction. In both comparisons, sensitivity was 0.50169 and specificity was 0.99724 where high precision is attributable to our applying the diagnostic predictive model to a large evaluation population. In ACEI vs ARB, incidence was 407/26792×100=1.52% and the OR changed from 1.18 to 1.23 (relative bias -4.23%). Similarly in ACEI vs THZ, incidence was 672/70158×100=0.96% and the OR changed from 1.09 to 1.13 (relative bias -3.62%). Under differential error correction in the same scenario, small outcome specificity differences between exposed populations led to a considerably larger OR for ACEI vs ARB (OR=1.38, relative bias=-17.7%) and an unstable result for ACEI vs THZ. Given incidence similarity between the cardiac defects study and our example, it is auspicious that phenotype errors from a conventional, internal validation study and those we produced using PheValuator were similar and resulted in similar bias correction. Chart adjudication-based validation studies, however, are time and resource intensive compared to a data-driven, probabilistic reference standard validation approach.

For ACEI vs ARB average bias reduction across 5 databases in the 730-day TAR, PS-matched vs PS-matched with non-differential QBA correction was 1.7% (range: 1.2-7.7%). Bias reduction from differential QBA correction was greater. Relative bias for the PS matched analysis with non-differential QBA correction vs differential QBA correction was almost as large as comparing no QBA correction to differential QBA correction (7.3% and 8.5%, respectively). This finding reinforces the importance of calculating exposure-specific outcome errors where possible because small specificity differences between exposure groups can strongly impact effect estimates[14, 26]. Although these small error differences can substantially impact bias, the many wide CIs and invalid results we observed in differential analyses may also be attributable to the lower outcome specificities in the exposed populations. That we observed unstable or invalid results in our empirical example, non-differential or differential notwithstanding, provides rationale for the synthetic grid space analysis. Alternatively, errors in sensitivity and specificity estimates could explain these findings.

Although we used PheValuator to support QBA methodological evaluation in pharmacoepidemiology, its value is worth discussion given the resource constraints on rapidly obtaining internal validation results for use with QBA (see **Web Appendix 5**).

## Synthetic grid space

Our preliminary empirical results showed invalid QBA inputs in certain scenarios, which informed our decision to conduct the synthetic grid space analysis to determine the space where QBA inputs are valid across pharmacoepidemiology scenarios. The impact of sensitivity was limited where incidence was low as relative effect estimates are not biased by non-differential sensitivity. As in the empirical example, small specificity changes had large impact on bias correction. Further, the synthetic analysis demonstrated that in scenarios with incidence of $10^{-1}$, QBA inputs are valid only where specificity is greater than 0.91. At lower incidence, the required valid input specificity increased with progressively greater precision. Lastly, the estimable proportion of scenarios decreased rapidly as we decreased incidence and increased the uncorrected estimates. This finding suggests that QBA for this purpose may be infeasible in common pharmacoepidemiology study settings where at least one input is incorrect.

## Strengths

First, despite minor data and design inconsistencies with the original study[20], our replicated estimates were of similar direction, magnitude, and precision. This provides some assurance that other sources of error were not responsible for the observed estimates before we applied QBA. Second, our empirical evaluation was a strictly systematic methodological assessment applied across multiple databases representing multiple real-world populations. Our assessment used data transformed to a CDM and standardized analytic tools that eliminated the possibility that variability in database structure or content, or analytic implementation could influence our results. Third, for researchers who have obtained exposure-agnostic outcome phenotype errors and have calculated its incidence in a study population, our synthetic grid space results represent a rough assessment on QBA feasibility across a range of possible effect estimates.

## Limitations

First, we did not evaluate probabilistic QBA, which accounts for phenotype error uncertainty. However, our multidimensional and synthetic analyses showed QBA impact across plausible sensitivity-specificity combinations. Further, our analyses assumed no selection bias or exposure or confounder misclassification. Second, our empirical evaluation assumed the phenotype errors we estimated were accurate. This is a non-trivial assumption considering QBA and similar methods are sensitive to measurement error accuracy[9, 27]. Third, we did not implement QBA using PPV and NPV. Doing so may be informative given most validation studies report PPV only and NPV is assumed high in low incidence scenarios[28]. We observed large, high-precision NPVs across databases, consistent with this assumption. Like specificity in low incidence settings, small deviations from perfect NPV will also have large impacts on estimates. Fourth, we used logistic regression outcome models, which facilitates QBA implementation and interpretation using 2x2 tables, but discards information in time-to-event data used in survival models commonly used in pharmacoepidemiology. This limitation warrants further QBA evaluation when applied to time-to-event data[12, 26]. The OR is a reasonable approximation of relative risk at low incidences we assessed, although this approximation could contribute to the larger relative bias results. Fifth, we created the synthetic data deterministically and did not include stochastic components. This constrained us from evaluating QBA impact on precision. Lastly, our QBA-corrected SE calculations assumed exposed populations independence, an assumption not fully met in our PS-matched analyses. This may result in inappropriately narrow CIs in our QBA-corrected PS-matched analyses and bias the relative precision metric.

## Conclusion

Our empirical and synthetic evaluation of QBA for outcome phenotype correction demonstrated that current implementations of QBA may be infeasible given erroneous or imprecise inputs in common pharmacoepidemiology scenarios. We also demonstrated that effect estimates are subject to considerable bias from outcome phenotype errors as the error magnitude increases. In other words, measurement error inaccuracy reduces the capacity for QBA to make adequate corrections. Regardless how phenotype errors are obtained, acknowledging them alone is unacceptable given the degree to which they can bias results on which clinical, policy, and regulatory decisions increasingly depend. Methods development for alternative approaches to phenotype error ascertainment and correction in low incidence settings warrants exploration.

# Tables

## Table 1: Exposure-agnostic and exposed population phenotype errors

| Database | Validation | Prevalence | TP | TN | FP | FN | Sensitivity | Specificity | PPV | NPV |
|---|---|---|---|---|---|---|---|---|---|---|
| Optum EHR® | Database | 4.923 | 35804 | 1820007 | 1007 | 58483 | 0.3797 | 0.9994 | 0.9726 | 0.9689 |
| Optum EHR® | ACEI | 12.627 | 88578 | 1245833 | 1565 | 91696 | 0.4914 | 0.9987 | 0.9826 | 0.9314 |
| Optum EHR® | ARB | 12.373 | 64130 | 947957 | 1552 | 69943 | 0.4783 | 0.9984 | 0.9764 | 0.9313 |
| Optum EHR® | THZ | 10.944 | 71205 | 1158355 | 1631 | 71347 | 0.4995 | 0.9986 | 0.9776 | 0.942 |
| Clinformatics® | Database | 3.662 | 41877 | 1864033 | 6157 | 29203 | 0.5892 | 0.9967 | 0.8718 | 0.9846 |
| Clinformatics® | ACEI | 9.567 | 78911 | 1188497 | 9747 | 47850 | 0.6225 | 0.9919 | 0.8901 | 0.9613 |
| Clinformatics® | ARB | 9.743 | 46228 | 725638 | 5460 | 32691 | 0.5858 | 0.9925 | 0.8944 | 0.9569 |
| Clinformatics® | THZ | 8.588 | 61054 | 1032787 | 7194 | 36652 | 0.6249 | 0.9931 | 0.8946 | 0.9657 |
| IBM CCAE | Database | 1.571 | 15309 | 1928187 | 2304 | 15503 | 0.4969 | 0.9988 | 0.8692 | 0.992 |
| IBM CCAE | ACEI | 5.902 | 39511 | 1219836 | 3518 | 37219 | 0.5149 | 0.9971 | 0.9182 | 0.9704 |
| IBM CCAE | ARB | 4.449 | 21487 | 751117 | 2615 | 13612 | 0.6122 | 0.9965 | 0.8915 | 0.9822 |
| IBM CCAE | THZ | 4.011 | 29844 | 1167007 | 4562 | 19109 | 0.6096 | 0.9961 | 0.8674 | 0.9839 |
| IBM MDCD | Database | 2.417 | 23184 | 1860538 | 5158 | 23028 | 0.5017 | 0.9972 | 0.818 | 0.9878 |
| IBM MDCD | ACEI | 10.282 | 20624 | 315886 | 3078 | 15932 | 0.5642 | 0.9904 | 0.8701 | 0.952 |
| IBM MDCD | ARB | 11.563 | 7485 | 106792 | 1007 | 6609 | 0.5311 | 0.9907 | 0.8814 | 0.9417 |
| IBM MDCD | THZ | 8.595 | 13651 | 255505 | 1907 | 10554 | 0.564 | 0.9926 | 0.8774 | 0.9603 |
| IBM MDCR | Database | 12.384 | 126037 | 1421229 | 14818 | 76940 | 0.6209 | 0.9897 | 0.8948 | 0.9486 |
| IBM MDCR | ACEI | 17.705 | 86120 | 622402 | 7646 | 49429 | 0.6353 | 0.9879 | 0.9185 | 0.9264 |
| IBM MDCR | ARB | 16.496 | 56294 | 440990 | 5502 | 31907 | 0.6382 | 0.9877 | 0.911 | 0.9325 |
| IBM MDCR | THZ | 15.105 | 71899 | 591685 | 7654 | 34740 | 0.6742 | 0.9872 | 0.9038 | 0.9445 |

Validation: population in which the ischemic stroke phenotype definition was validated against the probabilistic reference standard, Prevalence: estimated prevalence of ischemic stroke in the validation population (%) used to calibrate the PheValuator diagnostic predictive model, TP: true positives, TN: true negatives, FP: false positives, FN: false negatives, PPV: positive predictive value, NPV negative predictive value, Optum EHR®: Optum® de-identified Electronic Health Record Database; Clinformatics®: Optum's de-identified Clinformatics® Data Mart Database; IBM CCAE: IBM MarketScan® Commercial Database; IBM MDCD: IBM MarketScan® MultiState Medicaid Database; IBM MDCR: IBM MarketScan Medicare Supplemental and Coordination of Benefits Database. "Database" values in the Validation column refer to exposure-agnostic validation results. In exposed population validation, outcomes were assessed any time after drug initiation. Evaluation cohorts were sampled to 2,000,000 patients if they were larger at first. TP+TN+FP+FN values will be less than the total number of patients in the evaluation cohort because of study period and age restrictions.

## Table 2: Empirical evaluation results, 730-day time-at-risk

Empirical example QBA evaluation metrics for ACEI vs ARB and ACEI vs THZ exposure comparisons among patients with hypertension. For each exposure comparison, the 730-day time-at-risk analysis comparisons evaluation metrics are presented for each database and averaged across databases. Database-specific results are reported where the phenotype-error corrected 2x2 table includes no negative cell counts. In rows where the value in the Database column is one of the study data sources, the evaluation metric represents the analysis comparison in that database. For rows where the value of the Database column is "Average", the value represents the evaluation metric average computed across the N data sources that produced a valid OR estimated from a QBA-corrected 2x2 table that included no negative cells.

| T | C | Analysis comparison | Database | N | Relative Bias | Bias difference | Relative precision |
|---|---|---|---|---|---|---|---|
| ACEI | ARB | 730d, Unadjusted vs QBA | Optum EHR® | | -1.649 | -0.016 | 28.881 |
| ACEI | ARB | 730d, Unadjusted vs QBA | Clinformatics® | | -6.725 | -0.065 | 64.805 |
| ACEI | ARB | 730d, Unadjusted vs QBA | IBM CCAE | | -8.365 | -0.080 | 66.839 |
| ACEI | ARB | 730d, Unadjusted vs QBA | IBM MDCD | | 2.655 | 0.027 | 36.890 |
| ACEI | ARB | 730d, Unadjusted vs QBA | IBM MDCR | | -27.426 | -0.242 | 89.966 |
| ACEI | ARB | 730d, Unadjusted vs QBA | Average | 5 | -8.302 | -0.075 | 57.476 |
| ACEI | ARB | 730d, Unadjusted vs diff QBA | Optum EHR® | | -22.557 | -0.203 | 68.802 |
| ACEI | ARB | 730d, Unadjusted vs diff QBA | Clinformatics® | | -13.935 | -0.130 | 99.640 |
| ACEI | ARB | 730d, Unadjusted vs diff QBA | IBM CCAE | | | | |
| ACEI | ARB | 730d, Unadjusted vs diff QBA | IBM MDCD | | 31.753 | 0.382 | 89.656 |
| ACEI | ARB | 730d, Unadjusted vs diff QBA | IBM MDCR | | -67.578 | -0.516 | 96.768 |
| ACEI | ARB | 730d, Unadjusted vs diff QBA | Average | 4 | -18.079 | -0.117 | 88.717 |
| ACEI | ARB | 730d, QBA vs diff QBA | Optum EHR® | | -20.569 | -0.187 | 56.133 |
| ACEI | ARB | 730d, QBA vs diff QBA | Clinformatics® | | -6.756 | -0.065 | 98.978 |
| ACEI | ARB | 730d, QBA vs diff QBA | IBM CCAE | | | | |
| ACEI | ARB | 730d, QBA vs diff QBA | IBM MDCD | | 29.892 | 0.355 | 83.610 |
| ACEI | ARB | 730d, QBA vs diff QBA | IBM MDCR | | -31.511 | -0.274 | 67.792 |
| ACEI | ARB | 730d, QBA vs diff QBA | Average | 4 | -7.236 | -0.043 | 76.628 |
| ACEI | ARB | 730d, PS matched vs PS matched, QBA | Optum EHR® | | -1.221 | -0.012 | 29.484 |
| ACEI | ARB | 730d, PS matched vs PS matched, QBA | Clinformatics® | | -6.343 | -0.062 | 63.993 |
| ACEI | ARB | 730d, PS matched vs PS matched, QBA | IBM CCAE | | -4.354 | -0.043 | 66.658 |
| ACEI | ARB | 730d, PS matched vs PS matched, QBA | IBM MDCD | | -4.231 | -0.041 | 35.832 |
| ACEI | ARB | 730d, PS matched vs PS matched, QBA | IBM MDCR | | 7.671 | 0.080 | 91.085 |
| ACEI | ARB | 730d, PS matched vs PS matched, QBA | Average | 5 | -1.696 | -0.016 | 57.410 |
| ACEI | ARB | 730d, PS matched vs PS matched, diff QBA | Optum EHR® | | -21.201 | -0.192 | 65.956 |
| ACEI | ARB | 730d, PS matched vs PS matched, diff QBA | Clinformatics® | | -3.369 | -0.033 | 99.612 |
| ACEI | ARB | 730d, PS matched vs PS matched, diff QBA | IBM CCAE | | | | |
| ACEI | ARB | 730d, PS matched vs PS matched, diff QBA | IBM MDCD | | -17.692 | -0.163 | 86.985 |
| ACEI | ARB | 730d, PS matched vs PS matched, diff QBA | IBM MDCR | | 8.274 | 0.086 | 97.113 |
| ACEI | ARB | 730d, PS matched vs PS matched, diff QBA | Average | 4 | -8.497 | -0.075 | 87.417 |
| ACEI | ARB | 730d, PS matched, QBA vs PS matched, diff QBA | Optum EHR® | | -19.739 | -0.180 | 51.722 |
| ACEI | ARB | 730d, PS matched, QBA vs PS matched, diff QBA | Clinformatics® | | 2.797 | 0.028 | 98.922 |
| ACEI | ARB | 730d, PS matched, QBA vs PS matched, diff QBA | IBM CCAE | | | | |
| ACEI | ARB | 730d, PS matched, QBA vs PS matched, diff QBA | IBM MDCD | | -12.914 | -0.121 | 79.717 |
| ACEI | ARB | 730d, PS matched, QBA vs PS matched, diff QBA | IBM MDCR | | 0.653 | 0.007 | 67.622 |
| ACEI | ARB | 730d, PS matched, QBA vs PS matched, diff QBA | Average | 4 | -7.301 | -0.067 | 74.496 |
| ACEI | THZ | 730d, Unadjusted vs QBA | Optum EHR® | | -8.342 | -0.080 | 36.041 |
| ACEI | THZ | 730d, Unadjusted vs QBA | Clinformatics® | | -21.345 | -0.193 | 71.722 |
| ACEI | THZ | 730d, Unadjusted vs QBA | IBM CCAE | | -11.666 | -0.110 | 68.237 |
| ACEI | THZ | 730d, Unadjusted vs QBA | IBM MDCD | | -19.612 | -0.179 | 53.606 |
| ACEI | THZ | 730d, Unadjusted vs QBA | IBM MDCR | | -19.853 | -0.181 | 89.031 |
| ACEI | THZ | 730d, Unadjusted vs QBA | Average | 5 | -16.164 | -0.149 | 63.727 |
| ACEI | THZ | 730d, Unadjusted vs diff QBA | Optum EHR® | | -45.704 | -0.376 | 75.631 |
| ACEI | THZ | 730d, Unadjusted vs diff QBA | Clinformatics® | | | | |
| ACEI | THZ | 730d, Unadjusted vs diff QBA | IBM CCAE | | | | |
| ACEI | THZ | 730d, Unadjusted vs diff QBA | IBM MDCD | | -188.590 | -1.060 | 99.047 |
| ACEI | THZ | 730d, Unadjusted vs diff QBA | IBM MDCR | | -85.289 | -0.617 | 97.099 |
| ACEI | THZ | 730d, Unadjusted vs diff QBA | Average | 3 | -106.528 | -0.684 | 90.592 |
| ACEI | THZ | 730d, QBA vs diff QBA | Optum EHR® | | -34.485 | -0.296 | 61.899 |
| ACEI | THZ | 730d, QBA vs diff QBA | Clinformatics® | | | | |
| ACEI | THZ | 730d, QBA vs diff QBA | IBM CCAE | | | | |
| ACEI | THZ | 730d, QBA vs diff QBA | IBM MDCD | | -141.272 | -0.881 | 97.945 |
| ACEI | THZ | 730d, QBA vs diff QBA | IBM MDCR | | -54.597 | -0.436 | 73.554 |
| ACEI | THZ | 730d, QBA vs diff QBA | Average | 3 | -76.784 | -0.538 | 77.799 |
| ACEI | THZ | 730d, PS matched vs PS matched, QBA | Optum EHR® | | -3.900 | -0.038 | 35.144 |
| ACEI | THZ | 730d, PS matched vs PS matched, QBA | Clinformatics® | | -9.848 | -0.094 | 72.310 |
| ACEI | THZ | 730d, PS matched vs PS matched, QBA | IBM CCAE | | 6.965 | 0.072 | 71.258 |
| ACEI | THZ | 730d, PS matched vs PS matched, QBA | IBM MDCD | | -3.616 | -0.036 | 50.411 |
| ACEI | THZ | 730d, PS matched vs PS matched, QBA | IBM MDCR | | 7.464 | 0.078 | 91.581 |
| ACEI | THZ | 730d, PS matched vs PS matched, QBA | Average | 5 | -0.587 | -0.004 | 64.141 |
| ACEI | THZ | 730d, PS matched vs PS matched, diff QBA | Optum EHR® | | -28.074 | -0.247 | 72.321 |
| ACEI | THZ | 730d, PS matched vs PS matched, diff QBA | Clinformatics® | | | | |
| ACEI | THZ | 730d, PS matched vs PS matched, diff QBA | IBM CCAE | | | | |
| ACEI | THZ | 730d, PS matched vs PS matched, diff QBA | IBM MDCD | | 82.969 | 1.770 | 99.780 |
| ACEI | THZ | 730d, PS matched vs PS matched, diff QBA | IBM MDCR | | -19.262 | -0.176 | 97.916 |
| ACEI | THZ | 730d, PS matched vs PS matched, diff QBA | Average | 3 | 11.878 | 0.449 | 90.006 |
| ACEI | THZ | 730d, PS matched, QBA vs PS matched, diff QBA | Optum EHR® | | -23.266 | -0.209 | 57.323 |
| ACEI | THZ | 730d, PS matched, QBA vs PS matched, diff QBA | Clinformatics® | | | | |
| ACEI | THZ | 730d, PS matched, QBA vs PS matched, diff QBA | IBM CCAE | | | | |
| ACEI | THZ | 730d, PS matched, QBA vs PS matched, diff QBA | IBM MDCD | | 83.563 | 1.806 | 99.556 |
| ACEI | THZ | 730d, PS matched, QBA vs PS matched, diff QBA | IBM MDCR | | -28.882 | -0.254 | 75.251 |
| ACEI | THZ | 730d, PS matched, QBA vs PS matched, diff QBA | Average | 3 | 10.472 | 0.448 | 77.377 |

T: target exposure; C: comparator exposure; ACEI: angiotensin-converting enzyme inhibitors; ARB: angiotensin receptor blockers; THZ: Thiazide/thiazide-like diuretics; d: day, QBA; quantitative bias analysis for non-differential outcome phenotype error correction; PS: propensity score; diff QBA: quantitative bias analysis for differential outcome phenotype error correction; missing cells indicate that QBA resulted in phenotype-error corrected 2x2 table that contained negative cell counts and the OR was non-estimable; Optum EHR®: Optum® de-identified Electronic Health Record Database; Clinformatics®: Optum's de-identified Clinformatics® Data Mart Database; IBM CCAE: IBM MarketScan® Commercial Database; IBM MDCD: IBM MarketScan® MultiState Medicaid Database; IBM MDCR: IBM MarketScan Medicare Supplemental and Coordination of Benefits Database

## Table 3: Synthetic grid space evaluation results

QBA evaluation metrics computed over the synthetic grid space of 12,000 scenarios comprised of 5 outcome incidence proportions [10-1, 10-2, 10-3, 10-4, 10-5] × 6 uncorrected odds ratios (OR) [1, 1.25, 1.50, 2, 4, 10] × 20 outcome sensitivity values [0.05 to 1.00 by 0.05] × 20 specificity values. Rows report evaluation metrics for the uncorrected OR vs the OR point estimate at the 50$^{th}$ percentile of the QBA-corrected distribution.

| IP | OR | $OR_{QBA}$ | Estimable | Sensitivity | Specificity | Bias difference | Relative bias |
|---|---|---|---|---|---|---|---|
| 0.1 | 1.001 | 1.002 | 0.855 | 0.6 | 0.947368 | -0.001 | -0.11 |
| 0.1 | 1.25 | 1.574 | 0.81 | 0.25 | 0.963158 | -0.230 | -25.89 |
| 0.1 | 1.5 | 2.152 | 0.72 | 0.4 | 0.957895 | -0.361 | -43.43 |
| 0.1 | 2 | 3.484 | 0.63 | 0.4 | 0.963158 | -0.555 | -74.22 |
| 0.1 | 4 | 8.445 | 0.3825 | 0.85 | 0.973684 | -0.747 | -111.12 |
| 0.1 | 10 | 24.858 | 0.2125 | 0.3 | 0.994737 | -0.911 | -148.58 |
| 0.01 | 1.001 | 1.002 | 0.95 | 0.5 | 0.995263 | -0.001 | -0.09 |
| 0.01 | 1.25 | 1.476 | 0.85 | 0.5 | 0.995789 | -0.166 | -18.06 |
| 0.01 | 1.5 | 2.048 | 0.8 | 1 | 0.995789 | -0.312 | -36.57 |
| 0.01 | 2 | 2.907 | 0.65 | 0.5 | 0.996842 | -0.374 | -45.33 |
| 0.01 | 4 | 7.252 | 0.4 | 1 | 0.997895 | -0.595 | -81.30 |
| 0.01 | 10 | 20.198 | 0.2 | 0.05 | 0.999474 | -0.703 | -101.98 |
| 0.001 | 1.001 | 1.000 | 0.95 | 0.525 | 0.999526 | 0.001 | 0.10 |
| 0.001 | 1.25 | 1.475 | 0.85 | 0.55 | 0.999579 | -0.166 | -18.00 |
| 0.001 | 1.5 | 1.943 | 0.8 | 0.05 | 0.999632 | -0.259 | -29.51 |
| 0.001 | 2 | 2.900 | 0.65 | 0.55 | 0.999684 | -0.371 | -44.99 |
| 0.001 | 4 | 6.104 | 0.4 | 0.05 | 0.999842 | -0.423 | -52.61 |
| 0.001 | 10 | 14.109 | 0.2 | 0.05 | 0.999947 | -0.344 | -41.09 |
| 0.0001 | 1.001 | 1.000 | 0.95 | 0.525 | 0.999953 | 0.001 | 0.10 |
| 0.0001 | 1.25 | 1.469 | 0.85 | 0.55 | 0.999958 | -0.162 | -17.54 |
| 0.0001 | 1.5 | 1.928 | 0.8 | 0.05 | 0.999963 | -0.251 | -28.56 |
| 0.0001 | 2 | 2.864 | 0.65 | 0.55 | 0.999968 | -0.359 | -43.18 |
| 0.0001 | 4 | 5.971 | 0.4 | 0.05 | 0.999984 | -0.401 | -49.27 |
| 0.0001 | 10 | 13.922 | 0.2 | 0.05 | 0.999995 | -0.331 | -39.22 |
| 0.00001 | 1.001 | 1.000 | 0.925 | 0.518243 | 0.999995 | 0.001 | 0.10 |
| 0.00001 | 1.25 | 1.418 | 0.83 | 0.75 | 0.999996 | -0.126 | -13.41 |
| 0.00001 | 1.5 | 1.927 | 0.735 | 0.7 | 0.999996 | -0.250 | -28.46 |
| 0.00001 | 2 | 2.562 | 0.64 | 0.65 | 0.999997 | -0.248 | -28.08 |
| 0.00001 | 4 | 5.957 | 0.355 | 0.5 | 0.999998 | -0.398 | -48.92 |
| 0.00001 | 10 | 11.858 | 0.165 | 0.4 | 0.999999 | -0.170 | -18.58 |

IP: incidence proportion input used to compute the synthetic 2x2 tables; OR: input odds ratio biased by phenotype error used to compute the synthetic 2x2 tables; QBA: quantitative bias analysis for outcome phenotype error correction; $OR_{QBA}$: output odds ratio corrected for phenotype error by QBA; Estimable: proportion of the 400 2x2 tables where a QBA-corrected OR could be computed, i.e., there were no negative cell counts in the corrected 2x2 table. Bias difference: the difference between the log of the QBA-corrected OR and the log of the input OR; Relative bias: difference between the QBA-corrected OR and the input OR divided by the input multiplied by 100.

# Figures

## Figure 1: Empirical example forest plots, 730d TAR

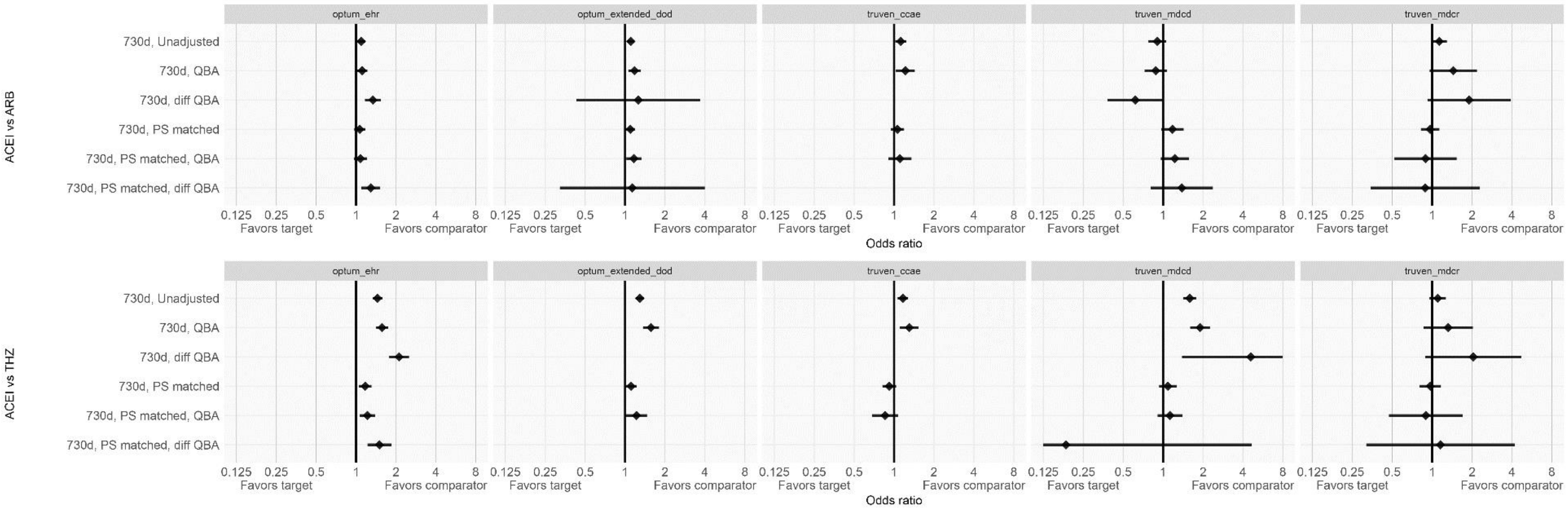

optum_ehr: Optum® de-identified Electronic Health Record Database; optum_extended_dod: Optum's de-identified Clinformatics® Data Mart Database; truven_ccae: IBM MarketScan® Commercial Database; truven_mdcd: IBM MarketScan® MultiState Medicaid Database; truven_mdcr: IBM MarketScan Medicare Supplemental and Coordination of Benefits Database

## Figure 2: Synthetic grid space results

QBA-corrected OR contour plots across synthetic grid space defined by incidence proportion (IP), uncorrected odds ratio (OR), sensitivity, and specificity. Plot rows represent IP strata and plot columns represent uncorrected OR strata. Within each IP-uncorrected OR stratum, x-axes represent sensitivity and y-axes represent specificity.

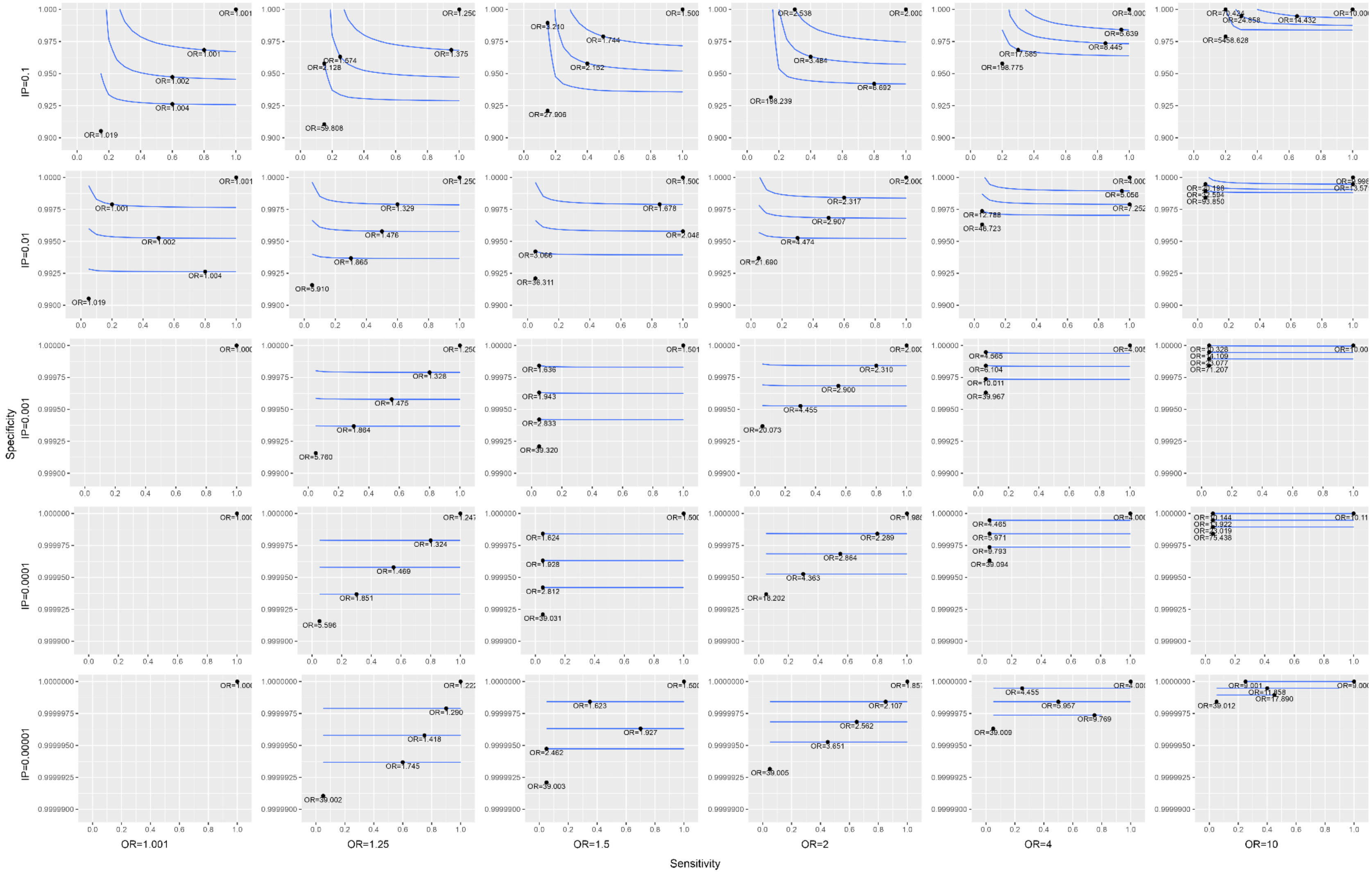

## Figure 3: Valid QBA input synthetic analytic space stratified by incidence and uncorrected odds ratio

Estimable proportion of QBA analyses for sensitivity by specificity combinations stratified by uncorrected odd ratio (OR) values and incidence proportion. Upper panel labels represent incidence proportion from $10^{-1}$ to $10^{-5}$. "Estimable" refers to the proportion of 400 synthetic 2x2 tables per stratum where QBA valid inputs produced corrected 2x2 tables that included all non-negative cell counts which allows computing a corrected OR.

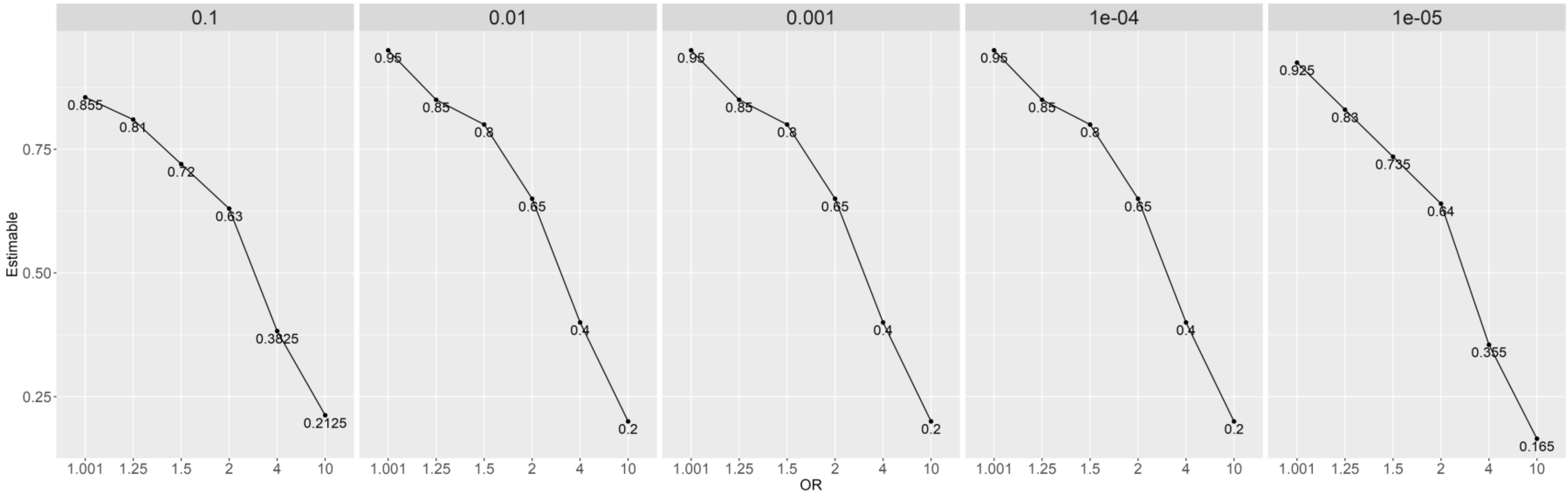

# Web Material

**Quantitative bias analysis for outcome phenotype error correction in comparative effect estimation: an empirical and synthetic evaluation**

James Weaver, Patrick B. Ryan, Victoria Y. Strauss, Marc A. Suchard, Joel Swerdel, Daniel Prieto-Alhambra

## Contents

## Web Table 1: Observed exposure-outcome 2x2 table

| | E1 | E0 |
|---|---|---|
| O[+] | a | b |
| O[-] | c | d |
| Total | $E1_{Total}$ | $E0_{Total}$ |

E1: Target, E0: Comparator, O[+]: Outcome, O[-]: No outcome

## Web Table 2: QBA-corrected exposure-outcome 2x2 table

| | E1 | E0 |
|---|---|---|
| O[+] | $A = [a-E1_{Total}*(1-SP_{E1})]/[(SE_{E1}-(1-SP_{E1})]$ | $B = [b-E0_{Total}*(1-SP_{E0})]/[(SE_{E0}-(1-SP_{E0})]$ |
| O[-] | $E1_{Total}-A$ | $E0_{Total}-B$ |
| Total | $E1_{Total}$ | $E0_{Total}$ |

E1: Target, E0: Comparator, O[+]: Outcome, O[-]: No outcome, $SN_{E1}$: Sensitivity in target, $SP_{E1}$: Specificity in target exposure population, $SN_{E0}$: Sensitivity in comparator, $SP_{E0}$: Specificity in comparator exposure population

## Web Appendix 1: Probabilistic reference standard validation

A PheValuator validation study requires at least 5 input specifications. First is the extremely specific cohort (xSpec), which narrowly identifies patients who are very likely a case. Second is the extremely sensitive cohort (xSens), which broadly identifies patients who could possibly be a case. Third, is the prevalence cohort, which is used as a crude measure of the phenotype prevalence in the database population. Fourth is the database evaluation cohort, which consists of patients with ≥1 occurrence of the outcome and patients with exactly 0 occurrences of the outcome selected from the full database population. In this study, we also used an additional evaluation cohort that is restricted to the exposed populations to calculate phenotype errors differential to exposure status. The exposed population evaluation cohorts consist of patients with ≥1 occurrence of the outcome with prior exposure and patients with exactly 0 occurrences of the outcome any time before or after an exposure. The detailed PheValuator cohort definitions are available in the **Protocol Appendix Section A** (https://ohdsi-studies.github.io/QbaEvaluation/protocol.html).

Our PheValuator validation study process was as follows.

1) Create a dataset of labeled (xSpec) and non-labeled (15,000 sampled database patients minus xSens patients) for building the diagnostic predictive model.
2) Build the diagnostic predictive model[1] to discriminate between labeled and non-labeled patients. Set the diagnostic predictive model intercept from the prevalence cohort, which will balance labeled and non-labeled patients to improve diagnostic predictive model efficiency. The diagnostic predictive model is roughly analogous to a clinical expert would perform chart adjudication is a traditional validation study.
3) Apply the diagnostic predictive model to the database evaluation cohort (2m sample) to produce a probability of outcome classification (discriminate O[+] from O[-]).
4) Apply the diagnostic predictive model to the exposed population evaluation cohorts to produce a probability of outcome classification among the exposed populations (discriminate exposed with O[+] from exposed with O[-]).
5) Populate the confusion matrix with conditional probabilities from the diagnostic predictive model stratified by phenotype definition classification and make measurement error calculations**.**

| | Phenotype algorithm under evaluation | |
|---|---|---|
| | Case ✓ | Non-case ✗ |
| Predicted outcome probability from the diagnostic model, P(Y) | TP = ∑ [P(Y\|Case)] | FP = ∑ [1 - P(Y\|Case)] |
| | FN = ∑ [P(Y\|Non-case)] | TN = ∑ [1 - P(Y\|Non-case)] |

TP: true positive, FP: false positive, FN: false negative, TN: true negative
Sensitivity=TP/(TP+FN), Specificity=TN/(TN+FP), PPV=TP/(TP+FP), NPV=TN/(TN+FN)

A strength of this approach is that it facilitates rapid, large-scale, multi-database, temporal, and study population-specific phenotype error ascertainment. Another strength is scalability. Because PheValuator uses a predictive model to assess potential cases, it can evaluate far more patient records than the small number usually reviewed by clinical experts who request, obtain, and adjudicate charts in a typical validation study. Time and resource constraints associated with manual chart review usually restrict adjudication activity to a random sample of patients who meet the phenotype definition criteria, which allows for calculating PPV. If the sample is large enough, the PPV may generalize to the database population from which the potential cases were sampled. To obtain NPV, sensitivity, and specificity, true and/or false negative counts are needed. Given low condition prevalence common to safety outcomes research, it is usually infeasible to review patient charts who do not meet the phenotype definition criteria to manually identify false negatives, since hundreds or thousands of records would need review to identify a single case misclassified as a non-case. PheValuator circumvents this problem by taking the sum of case probabilities among patients who do not qualify for the phenotype as the false negative count and taking the sum of 1 minus case probabilities among patients who do not qualify for the phenotype as the true negative count. A limitation of PheValuator is that it relies on information that has been translated from a clinical interaction to an electronic database, a multi-step process that is subject to missingness, error, and bias[2]. For example, many databases do not include unstructured clinical notes often available for manual review in patient charts. Further, discomfort with prediction as an acceptable alternative to clinical review for case ascertainment persists in the scientific community[3] despite evidence of performance similarity in some settings[4-6]

# Web Table 3: Database information

| Data source | Short name | Description |
|---|---|---|
| Optum® de-identified Electronic Health Record Dataset | Optum EHR® | Optum® de-identified Electronic Health Record Dataset is derived from dozens of healthcare provider organizations in the United States (that include more than 700 hospitals and 7,000 Clinics treating more than 103 million patients) receiving care in the United States. The medical record data includes clinical information, inclusive of prescriptions as prescribed and administered, lab results, vital signs, body measurements, diagnoses, procedures, and information derived from clinical Notes using Natural Language Processing (NLP). |
| Optum® de-Identified Clinformatics® Data Mart Database – Date of Death | Clinformatics® | Optum® De-Identified Clinformatics® Data Mart Database is an adjudicated administrative health claims database for members with private health insurance, who are fully insured in commercial plans or Medicare Advantage. The population is primarily representative of US commercial claims patients (0-65 years old) with some Medicare (65+ years old) however ages are capped at 90 years. It includes data captured from administrative claims processed from inpatient and outpatient medical services and prescriptions as dispensed, as well as results for outpatient lab tests processed by large national lab vendors who participate in data exchange with Optum. Optum DOD also provides date of death (month and year only) for members with both medical and pharmacy coverage from the Social Security Death Master File (however after 2011 reporting frequency changed due to changes in reporting requirements) and location information for patients is at the US state level. |
| IBM MarketScan® Commercial Claims and Encounters Database | IBM CCAE | IBM MarketScan® Commercial Claims and Encounters Database (CCAE) is a US employer-based private-payer administrative claims database. The data include adjudicated health insurance claims (e.g., inpatient, outpatient, and outpatient pharmacy) as well as enrollment data from large employers and health plans who provide private healthcare coverage to employees, their spouses, and dependents. Additionally, it captures laboratory tests for a subset of the covered lives. This administrative claims database includes a variety of fee-for-service, preferred provider organizations, and capitated health plans. |
| IBM MarketScan® MultiState Medicaid Database (MDCD) | IBM MDCD | IBM MarketScan® Multi-State Medicaid Database (MDCD) contains adjudicated US health insurance claims for Medicaid enrollees from multiple states and includes hospital discharge diagnoses, outpatient diagnoses and procedures, and outpatient pharmacy claims as well as ethnicity and Medicare eligibility. Members maintain their same identifier even if they leave the system for a brief period; however, the dataset lacks lab data. |
| IBM MarketScan® Medicare Supplemental and Coordination of Benefits Database | IBM MDCR | IBM MarketScan® Medicare Supplemental and Coordination of Benefits Database (MDCR) represents health services of retirees in the United States with primary or Medicare supplemental coverage through privately insured fee-for-service, point-of-service, or capitated health plans. These data include adjudicated health insurance claims (e.g., inpatient, outpatient, and outpatient pharmacy). Additionally, it captures laboratory tests for a subset of the covered lives. |

## Web Appendix 2: Web application instructions

All empirical and synthetic results are available in a web-based, interactive **Web Application** at https://data.ohdsi.org/QbaEvaluation/.

### Empirical example

By selecting a row in the main table in the Simple QBA tab, a user can review all effect estimates, exposure and outcome counts, patient attrition from design choices, population characteristics before and after PS-matching where applicable, the PS model, and PS and covariate balance diagnostics.

### Synthetic grid space

In the Synthetic grid space tab, selecting a value for incidence and observed OR allows a user to review all valid QBA results and evaluation metrics from all sensitivity-specificity combinations in an analytic stratum.

## Web Table 4: Empirical evaluation results, 365-day TAR

Empirical example QBA evaluation metrics for ACEI vs ARB and ACEI vs THZ exposure comparisons among patients with hypertension. For each exposure comparison, the 365-day time-at-risk analysis comparisons evaluation metrics are presented for each database and aggregated across databases as the average. Database-specific results are reported where the phenotype-error corrected 2x2 table includes no negative cell counts. In rows where the value in the Database column is one of the study data sources, the evaluation metric represents the analysis comparison in that database. For rows where the value of the Database column is "Average", the value represents the evaluation metric average computed across the N data sources that produced a valid OR estimated from a QBA-corrected 2x2 table that included no negative cells.

| T | C | Analysis comparison | Database | N | Relative Bias | Bias difference | Relative precision |
|---|---|---|---|---|---|---|---|
| ACEI | ARB | 365d, Unadjusted vs QBA | Optum EHR® | | -3.079 | -0.030 | 45.486 |
| ACEI | ARB | 365d, Unadjusted vs QBA | Clinformatics® | | -21.057 | -0.191 | 88.306 |
| ACEI | ARB | 365d, Unadjusted vs QBA | IBM CCAE | | -16.251 | -0.151 | 86.468 |
| ACEI | ARB | 365d, Unadjusted vs QBA | IBM MDCD | | 9.867 | 0.104 | 51.603 |
| ACEI | ARB | 365d, Unadjusted vs QBA | IBM MDCR | | | | |
| ACEI | ARB | 365d, Unadjusted vs QBA | Average | 4 | -7.630 | -0.067 | 67.966 |
| ACEI | ARB | 365d, Unadjusted vs diff QBA | Optum EHR® | | -100.451 | -0.695 | 94.235 |
| ACEI | ARB | 365d, Unadjusted vs diff QBA | Clinformatics® | | | | |
| ACEI | ARB | 365d, Unadjusted vs diff QBA | IBM CCAE | | | | |
| ACEI | ARB | 365d, Unadjusted vs diff QBA | IBM MDCD | | | | |
| ACEI | ARB | 365d, Unadjusted vs diff QBA | IBM MDCR | | | | |
| ACEI | ARB | 365d, Unadjusted vs diff QBA | Average | 1 | -100.451 | -0.695 | 94.235 |
| ACEI | ARB | 365d, QBA vs diff QBA | Optum EHR® | | -94.464 | -0.665 | 89.425 |
| ACEI | ARB | 365d, QBA vs diff QBA | Clinformatics® | | | | |
| ACEI | ARB | 365d, QBA vs diff QBA | IBM CCAE | | | | |
| ACEI | ARB | 365d, QBA vs diff QBA | IBM MDCD | | | | |
| ACEI | ARB | 365d, QBA vs diff QBA | IBM MDCR | | | | |
| ACEI | ARB | 365d, QBA vs diff QBA | Average | 1 | -94.464 | -0.665 | 89.425 |
| ACEI | ARB | 365d, PS matched vs PS matched, QBA | Optum EHR® | | -3.484 | -0.034 | 44.506 |
| ACEI | ARB | 365d, PS matched vs PS matched, QBA | Clinformatics® | | -23.942 | -0.215 | 86.824 |
| ACEI | ARB | 365d, PS matched vs PS matched, QBA | IBM CCAE | | -15.666 | -0.146 | 85.551 |
| ACEI | ARB | 365d, PS matched vs PS matched, QBA | IBM MDCD | | -4.171 | -0.041 | 49.556 |
| ACEI | ARB | 365d, PS matched vs PS matched, QBA | IBM MDCR | | | | |
| ACEI | ARB | 365d, PS matched vs PS matched, QBA | Average | 4 | -11.816 | -0.109 | 66.609 |
| ACEI | ARB | 365d, PS matched vs PS matched, diff QBA | Optum EHR® | | -105.505 | -0.720 | 92.620 |
| ACEI | ARB | 365d, PS matched vs PS matched, diff QBA | Clinformatics® | | | | |
| ACEI | ARB | 365d, PS matched vs PS matched, diff QBA | IBM CCAE | | | | |
| ACEI | ARB | 365d, PS matched vs PS matched, diff QBA | IBM MDCD | | | | |
| ACEI | ARB | 365d, PS matched vs PS matched, diff QBA | IBM MDCR | | | | |
| ACEI | ARB | 365d, PS matched vs PS matched, diff QBA | Average | 1 | -105.505 | -0.720 | 92.620 |
| ACEI | ARB | 365d, PS matched, QBA vs PS matched, diff QBA | Optum EHR® | | -98.586 | -0.686 | 86.700 |
| ACEI | ARB | 365d, PS matched, QBA vs PS matched, diff QBA | Clinformatics® | | | | |
| ACEI | ARB | 365d, PS matched, QBA vs PS matched, diff QBA | IBM CCAE | | | | |
| ACEI | ARB | 365d, PS matched, QBA vs PS matched, diff QBA | IBM MDCD | | | | |
| ACEI | ARB | 365d, PS matched, QBA vs PS matched, diff QBA | IBM MDCR | | | | |
| ACEI | ARB | 365d, PS matched, QBA vs PS matched, diff QBA | Average | 1 | -98.586 | -0.686 | 86.700 |
| ACEI | THZ | 365d, Unadjusted vs QBA | Optum EHR® | | -20.705 | -0.188 | 58.364 |
| ACEI | THZ | 365d, Unadjusted vs QBA | Clinformatics® | | -92.299 | -0.654 | 95.045 |
| ACEI | THZ | 365d, Unadjusted vs QBA | IBM CCAE | | -50.658 | -0.410 | 91.384 |
| ACEI | THZ | 365d, Unadjusted vs QBA | IBM MDCD | | -81.861 | -0.598 | 85.163 |
| ACEI | THZ | 365d, Unadjusted vs QBA | IBM MDCR | | | | |
| ACEI | THZ | 365d, Unadjusted vs QBA | Average | 4 | -61.381 | -0.462 | 82.489 |
| ACEI | THZ | 365d, Unadjusted vs diff QBA | Optum EHR® | | -708.315 | -2.090 | 99.612 |
| ACEI | THZ | 365d, Unadjusted vs diff QBA | Clinformatics® | | | | |
| ACEI | THZ | 365d, Unadjusted vs diff QBA | IBM CCAE | | | | |
| ACEI | THZ | 365d, Unadjusted vs diff QBA | IBM MDCD | | | | |
| ACEI | THZ | 365d, Unadjusted vs diff QBA | IBM MDCR | | | | |
| ACEI | THZ | 365d, Unadjusted vs diff QBA | Average | 1 | -708.315 | -2.090 | 99.612 |
| ACEI | THZ | 365d, QBA vs diff QBA | Optum EHR® | | -569.662 | -1.902 | 99.068 |
| ACEI | THZ | 365d, QBA vs diff QBA | Clinformatics® | | | | |
| ACEI | THZ | 365d, QBA vs diff QBA | IBM CCAE | | | | |
| ACEI | THZ | 365d, QBA vs diff QBA | IBM MDCD | | | | |
| ACEI | THZ | 365d, QBA vs diff QBA | IBM MDCR | | | | |
| ACEI | THZ | 365d, QBA vs diff QBA | Average | 1 | -569.662 | -1.902 | 99.068 |
| ACEI | THZ | 365d, PS matched vs PS matched, QBA | Optum EHR® | | -9.281 | -0.089 | 55.559 |
| ACEI | THZ | 365d, PS matched vs PS matched, QBA | Clinformatics® | | -48.067 | -0.392 | 95.663 |
| ACEI | THZ | 365d, PS matched vs PS matched, QBA | IBM CCAE | | -1.393 | -0.014 | 91.421 |
| ACEI | THZ | 365d, PS matched vs PS matched, QBA | IBM MDCD | | -25.261 | -0.225 | 76.755 |
| ACEI | THZ | 365d, PS matched vs PS matched, QBA | IBM MDCR | | | | |
| ACEI | THZ | 365d, PS matched vs PS matched, QBA | Average | 4 | -21.000 | -0.180 | 79.849 |
| ACEI | THZ | 365d, PS matched vs PS matched, diff QBA | Optum EHR® | | -258.508 | -1.277 | 98.547 |
| ACEI | THZ | 365d, PS matched vs PS matched, diff QBA | Clinformatics® | | | | |
| ACEI | THZ | 365d, PS matched vs PS matched, diff QBA | IBM CCAE | | | | |
| ACEI | THZ | 365d, PS matched vs PS matched, diff QBA | IBM MDCD | | | | |
| ACEI | THZ | 365d, PS matched vs PS matched, diff QBA | IBM MDCR | | | | |
| ACEI | THZ | 365d, PS matched vs PS matched, diff QBA | Average | 1 | -258.508 | -1.277 | 98.547 |
| ACEI | THZ | 365d, PS matched, QBA vs PS matched, diff QBA | Optum EHR® | | -228.060 | -1.188 | 96.730 |
| ACEI | THZ | 365d, PS matched, QBA vs PS matched, diff QBA | Clinformatics® | | | | |
| ACEI | THZ | 365d, PS matched, QBA vs PS matched, diff QBA | IBM CCAE | | | | |
| ACEI | THZ | 365d, PS matched, QBA vs PS matched, diff QBA | IBM MDCD | | | | |
| ACEI | THZ | 365d, PS matched, QBA vs PS matched, diff QBA | IBM MDCR | | | | |
| ACEI | THZ | 365d, PS matched, QBA vs PS matched, diff QBA | Average | 1 | -228.060 | -1.188 | 96.730 |

T: target exposure; C: comparator exposure; ACEI: angiotensin-converting enzyme inhibitors; ARB: angiotensin receptor blockers; THZ: Thiazide/thiazide-like diuretics; d: day, QBA; quantitative bias analysis for non-differential outcome phenotype error correction; PS: propensity score; diff QBA: quantitative bias analysis for differential outcome phenotype error correction; missing cells indicate that QBA resulted in phenotype-error corrected 2x2 table that contained negative cell counts and the OR was non-estimable; Optum EHR®: Optum® de-identified Electronic Health Record Database; Clinformatics®: Optum's de-identified Clinformatics® Data Mart Database; IBM CCAE: IBM MarketScan® Commercial Database; IBM MDCD: IBM MarketScan® MultiState Medicaid Database; IBM MDCR: IBM MarketScan Medicare Supplemental and Coordination of Benefits Database

## Web Appendix 3: Multidimensional QBA results

Minimally varying specificity in multidimensional QBA had a large impact on the magnitude and precision of phenotype-error correction. For example, for ACEI vs ARB in Optum EHR® at fixed sensitivity of 0.589, decreasing the specificity from 0.997 to 0.9953 increased the uncorrected OR by 189%. Further decreasing the sensitivity to 0.995 invalidated the results by producing negative cells in the corrected 2x2 table counts (See Non-differential multidimensional QBA tab in the **Web Application**). The large impact of specificity precision is the result of low incidence scenarios. In **Web Tables 1 and 2**, note how cell *A* is corrected as $[a\text{-}E1_{Total}\times(1\text{-}SP_{E1})]/[(SE_{E1}\text{-}(1\text{-}SP_{E1})]$. In a scenario where $E1_{Total}$=100,000, $a$=100 (incidence=0.001), non-differential sensitivity=0.5 and specificity=1 the term $E1_{Total}\times(1\text{-}SP_{E1})$ reduces to 0, so the numerator of the equation equals $a$. But if we reduce specificity to 0.99 the term $E1_{Total}\times(1\text{-}SP_{E1})$ becomes 100,000×0.01=1,000, which when subtracted from 100 equals -900. Including this negative value in the cell *A* equation results in *A*=-1,837, which is an invalid result. For *A* to be positive, we require that $E1_{Total}\times(1\text{-}SP_{E1})<a$ which occurs at specificity=0.999004. In scenarios where $a$ is a very small proportion of $E1_{Total}$ extremely high specificity values are required for $E1_{Total}\times(1\text{-}SP_{E1})<a$ and small specificity precision changes have a large impact on corrected cell counts.

## Web Figure 1: Empirical example forest plots, 365d TAR

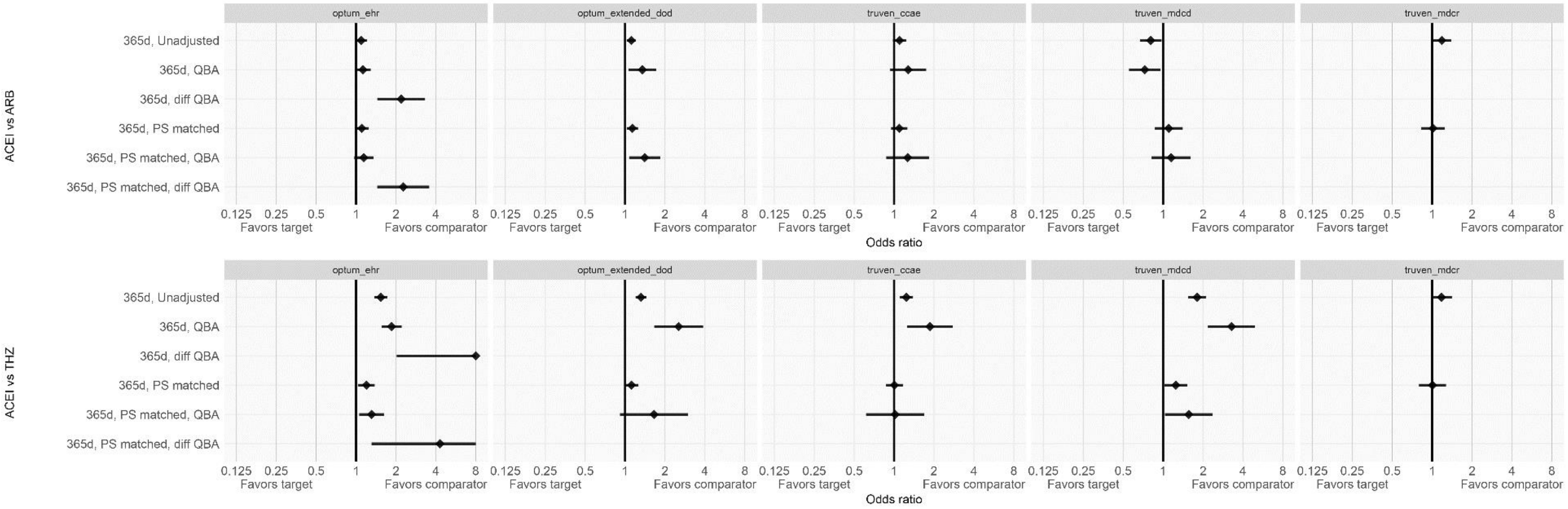

optum_ehr: Optum® de-identified Electronic Health Record Database; optum_extended_dod: Optum's de-identified Clinformatics® Data Mart Database; truven_ccae: IBM MarketScan® Commercial Database; truven_mdcd: IBM MarketScan® MultiState Medicaid Database; truven_mdcr: IBM MarketScan Medicare Supplemental and Coordination of Benefits Database

## Web Appendix 4: Synthetic grid space evaluation results

**Web Table 4** reports the same content but includes comparisons between the uncorrected OR and the 25th percentile, 75th percentile, and maximum value of the corrected OR distribution within each incidence-corrected OR strata. As expected, the magnitude of bias correction increases as the uncorrected OR is compared to progressively larger values of the corrected OR distribution. Compared to the 75% percentile of the corrected OR distribution where incidence is 0.1, relative bias correction is -6.04% at the highest uncorrected OR of 10 (corrected OR=70.42%). At this incidence, relative bias correction ranges implausibly from -46.8% to -544.9% when comparing uncorrected ORs of 1.25 or greater to the maximum valid corrected OR. For example, where the uncorrected OR is 2, the maximum valid corrected OR is 198 at 0.15 sensitivity and 0.93 specificity with a relative bias correction of -98.1%. Bias difference evaluation results are directionally consistent with relative bias evaluation results.

## Web Table 4: Synthetic grid space evaluation results

QBA evaluation metrics computed over the synthetic grid space of 12,000 scenarios comprised of 5 outcome incidence proportions [$10^{-1}$, $10^{-2}$, $10^{-3}$, $10^{-4}$, $10^{-5}$] × 6 uncorrected odds ratios (OR) [1, 1.25, 1.50, 2, 4, 10] × 20 outcome sensitivity values [0.05 to 1.00 by 0.05] × 20 specificity values. Rows report evaluation metrics for the uncorrected OR vs the OR point estimate at the minimum, 25th, 50th, 75th percentile, and maximum of the QBA-corrected distribution.

| Incidence | OR | Dist. | $OR_{QBA}$ | Estimable | Sensitivity | Specificity | Bias difference | Relative bias |
|---|---|---|---|---|---|---|---|---|
| 0.1 | 1.001 | min | 1.001 | 0.855 | 1 | 1 | 0.000 | 0.00 |
| 0.1 | 1.001 | 25%ile | 1.001 | 0.855 | 0.8 | 0.968421 | 0.000 | -0.04 |
| 0.1 | 1.001 | 50%ile | 1.002 | 0.855 | 0.6 | 0.947368 | -0.001 | -0.11 |
| 0.1 | 1.001 | 75%ile | 1.004 | 0.855 | 0.6 | 0.926316 | -0.003 | -0.26 |
| 0.1 | 1.001 | max | 1.019 | 0.855 | 0.15 | 0.905263 | -0.018 | -1.81 |
| 0.1 | 1.25 | min | 1.250 | 0.81 | 1 | 1 | 0.000 | 0.00 |
| 0.1 | 1.25 | 25%ile | 1.375 | 0.81 | 0.95 | 0.968421 | -0.095 | -9.98 |
| 0.1 | 1.25 | 50%ile | 1.574 | 0.81 | 0.25 | 0.963158 | -0.230 | -25.89 |
| 0.1 | 1.25 | 75%ile | 2.128 | 0.81 | 0.15 | 0.957895 | -0.532 | -70.25 |
| 0.1 | 1.25 | max | 59.808 | 0.81 | 0.15 | 0.910526 | -3.868 | -4684.65 |
| 0.1 | 1.5 | min | 1.500 | 0.72 | 1 | 1 | 0.000 | 0.00 |
| 0.1 | 1.5 | 25%ile | 1.744 | 0.72 | 0.5 | 0.978947 | -0.151 | -16.28 |
| 0.1 | 1.5 | 50%ile | 2.152 | 0.72 | 0.4 | 0.957895 | -0.361 | -43.43 |
| 0.1 | 1.5 | 75%ile | 3.210 | 0.72 | 0.15 | 0.989474 | -0.761 | -113.98 |
| 0.1 | 1.5 | max | 27.906 | 0.72 | 0.15 | 0.921053 | -2.923 | -1760.42 |
| 0.1 | 2 | min | 2.000 | 0.63 | 1 | 1 | 0.000 | 0.00 |
| 0.1 | 2 | 25%ile | 2.538 | 0.63 | 0.3 | 1 | -0.238 | -26.88 |
| 0.1 | 2 | 50%ile | 3.484 | 0.63 | 0.4 | 0.963158 | -0.555 | -74.22 |
| 0.1 | 2 | 75%ile | 6.692 | 0.63 | 0.8 | 0.942105 | -1.208 | -234.62 |
| 0.1 | 2 | max | 198.239 | 0.63 | 0.15 | 0.931579 | -4.596 | -9811.94 |
| 0.1 | 4 | min | 4.000 | 0.3825 | 1 | 1 | 0.000 | 0.00 |
| 0.1 | 4 | 25%ile | 5.639 | 0.3825 | 0.95 | 0.984211 | -0.343 | -40.98 |
| 0.1 | 4 | 50%ile | 8.445 | 0.3825 | 0.85 | 0.973684 | -0.747 | -111.12 |
| 0.1 | 4 | 75%ile | 17.585 | 0.3825 | 0.3 | 0.968421 | -1.481 | -339.64 |
| 0.1 | 4 | max | 198.775 | 0.3825 | 0.2 | 0.957895 | -3.906 | -4869.37 |
| 0.1 | 10 | min | 10.000 | 0.2125 | 1 | 1 | 0.000 | 0.00 |
| 0.1 | 10 | 25%ile | 14.432 | 0.2125 | 0.65 | 0.994737 | -0.367 | -44.32 |
| 0.1 | 10 | 50%ile | 24.858 | 0.2125 | 0.3 | 0.994737 | -0.911 | -148.58 |
| 0.1 | 10 | 75%ile | 70.424 | 0.2125 | 0.2 | 1 | -1.952 | -604.24 |
| 0.1 | 10 | max | 5458.628 | 0.2125 | 0.2 | 0.978947 | -6.302 | -54486.28 |
| 0.01 | 1.001 | min | 1.001 | 0.95 | 1 | 1 | 0.000 | 0.00 |
| 0.01 | 1.001 | 25%ile | 1.001 | 0.95 | 0.2 | 0.997895 | 0.000 | -0.03 |
| 0.01 | 1.001 | 50%ile | 1.002 | 0.95 | 0.5 | 0.995263 | -0.001 | -0.09 |
| 0.01 | 1.001 | 75%ile | 1.004 | 0.95 | 0.8 | 0.992632 | -0.003 | -0.28 |
| 0.01 | 1.001 | max | 1.019 | 0.95 | 0.05 | 0.990526 | -0.018 | -1.84 |
| 0.01 | 1.25 | min | 1.250 | 0.85 | 1 | 1 | 0.000 | 0.00 |
| 0.01 | 1.25 | 25%ile | 1.329 | 0.85 | 0.6 | 0.997895 | -0.061 | -6.30 |
| 0.01 | 1.25 | 50%ile | 1.476 | 0.85 | 0.5 | 0.995789 | -0.166 | -18.06 |
| 0.01 | 1.25 | 75%ile | 1.865 | 0.85 | 0.3 | 0.993684 | -0.400 | -49.23 |
| 0.01 | 1.25 | max | 5.910 | 0.85 | 0.05 | 0.991579 | -1.553 | -372.78 |
| 0.01 | 1.5 | min | 1.500 | 0.8 | 1 | 1 | 0.000 | 0.00 |
| 0.01 | 1.5 | 25%ile | 1.678 | 0.8 | 0.85 | 0.997895 | -0.112 | -11.86 |
| 0.01 | 1.5 | 50%ile | 2.048 | 0.8 | 1 | 0.995789 | -0.312 | -36.57 |
| 0.01 | 1.5 | 75%ile | 3.066 | 0.8 | 0.05 | 0.994211 | -0.715 | -104.43 |
| 0.01 | 1.5 | max | 36.311 | 0.8 | 0.05 | 0.992105 | -3.187 | -2320.72 |
| 0.01 | 2 | min | 2.000 | 0.65 | 1 | 1 | 0.000 | -0.01 |
| 0.01 | 2 | 25%ile | 2.317 | 0.65 | 0.6 | 0.998421 | -0.147 | -15.86 |
| 0.01 | 2 | 50%ile | 2.907 | 0.65 | 0.5 | 0.996842 | -0.374 | -45.33 |
| 0.01 | 2 | 75%ile | 4.474 | 0.65 | 0.3 | 0.995263 | -0.805 | -123.68 |
| 0.01 | 2 | max | 21.690 | 0.65 | 0.05 | 0.993684 | -2.384 | -984.48 |
| 0.01 | 4 | min | 4.000 | 0.4 | 1 | 1 | 0.000 | 0.01 |
| 0.01 | 4 | 25%ile | 5.056 | 0.4 | 0.95 | 0.998947 | -0.234 | -26.40 |
| 0.01 | 4 | 50%ile | 7.252 | 0.4 | 1 | 0.997895 | -0.595 | -81.30 |
| 0.01 | 4 | 75%ile | 12.788 | 0.4 | 0.05 | 0.997368 | -1.162 | -219.70 |
| 0.01 | 4 | max | 46.723 | 0.4 | 0.05 | 0.996316 | -2.458 | -1068.06 |
| 0.01 | 10 | min | 9.998 | 0.2 | 1 | 1 | 0.000 | 0.02 |
| 0.01 | 10 | 25%ile | 13.579 | 0.2 | 1 | 0.999474 | -0.306 | -35.79 |
| 0.01 | 10 | 50%ile | 20.198 | 0.2 | 0.05 | 0.999474 | -0.703 | -101.98 |
| 0.01 | 10 | 75%ile | 32.594 | 0.2 | 0.05 | 0.998947 | -1.182 | -225.94 |
| 0.01 | 10 | max | 93.850 | 0.2 | 0.05 | 0.998421 | -2.239 | -838.50 |
| 0.001 | 1.001 | min | 1.000 | 0.95 | 0.525 | 0.999526 | 0.001 | 0.10 |
| 0.001 | 1.001 | 25%ile | 1.000 | 0.95 | 0.525 | 0.999526 | 0.001 | 0.10 |
| 0.001 | 1.001 | 50%ile | 1.000 | 0.95 | 0.525 | 0.999526 | 0.001 | 0.10 |
| 0.001 | 1.001 | 75%ile | 1.000 | 0.95 | 0.525 | 0.999526 | 0.001 | 0.10 |
| 0.001 | 1.001 | max | 1.000 | 0.95 | 0.525 | 0.999526 | 0.001 | 0.10 |
| 0.001 | 1.25 | min | 1.250 | 0.85 | 1 | 1 | 0.000 | 0.00 |
| 0.001 | 1.25 | 25%ile | 1.328 | 0.85 | 0.8 | 0.999789 | -0.060 | -6.21 |
| 0.001 | 1.25 | 50%ile | 1.475 | 0.85 | 0.55 | 0.999579 | -0.166 | -18.00 |
| 0.001 | 1.25 | 75%ile | 1.864 | 0.85 | 0.3 | 0.999368 | -0.399 | -49.10 |
| 0.001 | 1.25 | max | 5.760 | 0.85 | 0.05 | 0.999158 | -1.528 | -360.80 |
| 0.001 | 1.5 | min | 1.501 | 0.8 | 1 | 1 | 0.000 | -0.04 |
| 0.001 | 1.5 | 25%ile | 1.636 | 0.8 | 0.05 | 0.999842 | -0.087 | -9.08 |
| 0.001 | 1.5 | 50%ile | 1.943 | 0.8 | 0.05 | 0.999632 | -0.259 | -29.51 |
| 0.001 | 1.5 | 75%ile | 2.833 | 0.8 | 0.05 | 0.999421 | -0.636 | -88.84 |
| 0.001 | 1.5 | max | 39.320 | 0.8 | 0.05 | 0.999211 | -3.266 | -2521.31 |
| 0.001 | 2 | min | 2.000 | 0.65 | 1 | 1 | 0.000 | 0.01 |
| 0.001 | 2 | 25%ile | 2.310 | 0.65 | 0.8 | 0.999842 | -0.144 | -15.51 |
| 0.001 | 2 | 50%ile | 2.900 | 0.65 | 0.55 | 0.999684 | -0.371 | -44.99 |
| 0.001 | 2 | 75%ile | 4.455 | 0.65 | 0.3 | 0.999526 | -0.801 | -122.75 |
| 0.001 | 2 | max | 20.073 | 0.65 | 0.05 | 0.999368 | -2.306 | -903.67 |
| 0.001 | 4 | min | 4.005 | 0.4 | 1 | 1 | -0.001 | -0.12 |
| 0.001 | 4 | 25%ile | 4.565 | 0.4 | 0.05 | 0.999947 | -0.132 | -14.12 |
| 0.001 | 4 | 50%ile | 6.104 | 0.4 | 0.05 | 0.999842 | -0.423 | -52.61 |
| 0.001 | 4 | 75%ile | 10.011 | 0.4 | 0.05 | 0.999737 | -0.917 | -150.29 |
| 0.001 | 4 | max | 39.967 | 0.4 | 0.05 | 0.999632 | -2.302 | -899.17 |
| 0.001 | 10 | min | 10.005 | 0.2 | 1 | 1 | -0.001 | -0.05 |
| 0.001 | 10 | 25%ile | 10.328 | 0.2 | 0.05 | 1 | -0.032 | -3.28 |
| 0.001 | 10 | 50%ile | 14.109 | 0.2 | 0.05 | 0.999947 | -0.344 | -41.09 |
| 0.001 | 10 | 75%ile | 23.077 | 0.2 | 0.05 | 0.999895 | -0.836 | -130.77 |
| 0.001 | 10 | max | 71.207 | 0.2 | 0.05 | 0.999842 | -1.963 | -612.07 |
| 0.0001 | 1.001 | min | 1.000 | 0.95 | 0.525 | 0.999953 | 0.001 | 0.10 |
| 0.0001 | 1.001 | 25%ile | 1.000 | 0.95 | 0.525 | 0.999953 | 0.001 | 0.10 |
| 0.0001 | 1.001 | 50%ile | 1.000 | 0.95 | 0.525 | 0.999953 | 0.001 | 0.10 |
| 0.0001 | 1.001 | 75%ile | 1.000 | 0.95 | 0.525 | 0.999953 | 0.001 | 0.10 |
| 0.0001 | 1.001 | max | 1.000 | 0.95 | 0.525 | 0.999953 | 0.001 | 0.10 |
| 0.0001 | 1.25 | min | 1.247 | 0.85 | 1 | 1 | 0.002 | 0.22 |
| 0.0001 | 1.25 | 25%ile | 1.324 | 0.85 | 0.8 | 0.999979 | -0.057 | -5.91 |
| 0.0001 | 1.25 | 50%ile | 1.469 | 0.85 | 0.55 | 0.999958 | -0.162 | -17.54 |
| 0.0001 | 1.25 | 75%ile | 1.851 | 0.85 | 0.3 | 0.999937 | -0.393 | -48.12 |
| 0.0001 | 1.25 | max | 5.596 | 0.85 | 0.05 | 0.999916 | -1.499 | -347.67 |
| 0.0001 | 1.5 | min | 1.500 | 0.8 | 1 | 1 | 0.000 | 0.00 |
| 0.0001 | 1.5 | 25%ile | 1.624 | 0.8 | 0.05 | 0.999984 | -0.080 | -8.28 |
| 0.0001 | 1.5 | 50%ile | 1.928 | 0.8 | 0.05 | 0.999963 | -0.251 | -28.56 |
| 0.0001 | 1.5 | 75%ile | 2.812 | 0.8 | 0.05 | 0.999942 | -0.628 | -87.45 |
| 0.0001 | 1.5 | max | 39.031 | 0.8 | 0.05 | 0.999921 | -3.259 | -2502.09 |
| 0.0001 | 2 | min | 1.985 | 0.65 | 1 | 1 | 0.007 | 0.74 |
| 0.0001 | 2 | 25%ile | 2.289 | 0.65 | 0.8 | 0.999984 | -0.135 | -14.45 |
| 0.0001 | 2 | 50%ile | 2.864 | 0.65 | 0.55 | 0.999968 | -0.359 | -43.18 |
| 0.0001 | 2 | 75%ile | 4.363 | 0.65 | 0.3 | 0.999953 | -0.780 | -118.14 |
| 0.0001 | 2 | max | 18.202 | 0.65 | 0.05 | 0.999937 | -2.208 | -810.11 |
| 0.0001 | 4 | min | 4.000 | 0.4 | 1 | 1 | 0.000 | -0.01 |
| 0.0001 | 4 | 25%ile | 4.465 | 0.4 | 0.05 | 0.999995 | -0.110 | -11.63 |
| 0.0001 | 4 | 50%ile | 5.971 | 0.4 | 0.05 | 0.999984 | -0.401 | -49.27 |
| 0.0001 | 4 | 75%ile | 9.793 | 0.4 | 0.05 | 0.999974 | -0.895 | -144.82 |
| 0.0001 | 4 | max | 39.094 | 0.4 | 0.05 | 0.999963 | -2.280 | -877.35 |
| 0.0001 | 10 | min | 10.113 | 0.2 | 1 | 1 | -0.011 | -1.13 |
| 0.0001 | 10 | 25%ile | 10.144 | 0.2 | 0.05 | 1 | -0.014 | -1.44 |
| 0.0001 | 10 | 50%ile | 13.922 | 0.2 | 0.05 | 0.999995 | -0.331 | -39.22 |
| 0.0001 | 10 | 75%ile | 23.019 | 0.2 | 0.05 | 0.999989 | -0.834 | -130.19 |
| 0.0001 | 10 | max | 75.438 | 0.2 | 0.05 | 0.999984 | -2.021 | -654.38 |
| 0.00001 | 1.001 | min | 1.000 | 0.925 | 0.518243 | 0.999995 | 0.001 | 0.10 |
| 0.00001 | 1.001 | 25%ile | 1.000 | 0.925 | 0.518243 | 0.999995 | 0.001 | 0.10 |
| 0.00001 | 1.001 | 50%ile | 1.000 | 0.925 | 0.518243 | 0.999995 | 0.001 | 0.10 |
| 0.00001 | 1.001 | 75%ile | 1.000 | 0.925 | 0.518243 | 0.999995 | 0.001 | 0.10 |
| 0.00001 | 1.001 | max | 1.000 | 0.925 | 0.518243 | 0.999995 | 0.001 | 0.10 |
| 0.00001 | 1.25 | min | 1.222 | 0.83 | 1 | 1 | 0.022 | 2.22 |
| 0.00001 | 1.25 | 25%ile | 1.290 | 0.83 | 0.9 | 0.999998 | -0.032 | -3.21 |
| 0.00001 | 1.25 | 50%ile | 1.418 | 0.83 | 0.75 | 0.999996 | -0.126 | -13.41 |
| 0.00001 | 1.25 | 75%ile | 1.745 | 0.83 | 0.6 | 0.999994 | -0.334 | -39.61 |
| 0.00001 | 1.25 | max | 39.002 | 0.83 | 0.05 | 0.999991 | -3.440 | -3020.12 |
| 0.00001 | 1.5 | min | 1.500 | 0.735 | 1 | 1 | 0.000 | 0.00 |
| 0.00001 | 1.5 | 25%ile | 1.623 | 0.735 | 0.35 | 0.999998 | -0.079 | -8.20 |
| 0.00001 | 1.5 | 50%ile | 1.927 | 0.735 | 0.7 | 0.999996 | -0.250 | -28.46 |
| 0.00001 | 1.5 | 75%ile | 2.462 | 0.735 | 0.05 | 0.999995 | -0.495 | -64.12 |
| 0.00001 | 1.5 | max | 39.003 | 0.735 | 0.05 | 0.999992 | -3.258 | -2500.21 |
| 0.00001 | 2 | min | 1.857 | 0.64 | 1 | 1 | 0.074 | 7.14 |
| 0.00001 | 2 | 25%ile | 2.107 | 0.64 | 0.85 | 0.999998 | -0.052 | -5.34 |
| 0.00001 | 2 | 50%ile | 2.562 | 0.64 | 0.65 | 0.999997 | -0.248 | -28.08 |
| 0.00001 | 2 | 75%ile | 3.651 | 0.64 | 0.45 | 0.999995 | -0.602 | -82.56 |
| 0.00001 | 2 | max | 39.005 | 0.64 | 0.05 | 0.999993 | -2.971 | -1850.23 |
| 0.00001 | 4 | min | 4.000 | 0.355 | 1 | 1 | 0.000 | 0.00 |
| 0.00001 | 4 | 25%ile | 4.455 | 0.355 | 0.25 | 0.999999 | -0.108 | -11.37 |
| 0.00001 | 4 | 50%ile | 5.957 | 0.355 | 0.5 | 0.999998 | -0.398 | -48.92 |
| 0.00001 | 4 | 75%ile | 9.769 | 0.355 | 0.75 | 0.999997 | -0.893 | -144.23 |
| 0.00001 | 4 | max | 39.009 | 0.355 | 0.05 | 0.999996 | -2.278 | -875.23 |
| 0.00001 | 10 | min | 9.000 | 0.165 | 1 | 1 | 0.105 | 10.00 |
| 0.00001 | 10 | 25%ile | 9.001 | 0.165 | 0.25 | 1 | 0.105 | 9.99 |
| 0.00001 | 10 | 50%ile | 11.858 | 0.165 | 0.4 | 0.999999 | -0.170 | -18.58 |
| 0.00001 | 10 | 75%ile | 17.890 | 0.165 | 0.45 | 0.999999 | -0.582 | -78.90 |
| 0.00001 | 10 | max | 39.012 | 0.165 | 0.05 | 0.999998 | -1.361 | -290.12 |

Incidence: incidence proportion input used to compute the synthetic 2x2 tables; OR: uncorrected odds ratio used is input to compute the synthetic 2x2 tables; QBA: quantitative bias analysis for outcome phenotype error correction; $OR_{QBA}$: output odds ratio corrected for phenotype error by QBA; Estimable: proportion of the 400 2x2 tables where a QBA-corrected OR could be computed, i.e., there were no negative cells in the corrected 2x2 table. Bias difference: the difference between the log of the QBA-corrected OR and the log of the input OR; Relative bias: difference between the QBA-corrected OR and the input OR divided by the input multiplied by 100.

## Web Appendix 5: Value of PheValuator

PheValuator is an emerging tool that overcomes the known problem of calculating sensitivity, specificity, and NPV in conventional validation studies[7]. To date, it has been used in clinical applications for estimating exposure-agnostic phenotype errors pulmonary hypertension, chronic thromboembolic pulmonary hypertension, systemic lupus erythematosus, and hidradenitis suppurativa [8-10]. PheValuator was evaluated by assessing its performance in estimating PPV compared to gold-standard validation studies across 90 definitions for 19 phenotypes in five therapeutic areas in five databases. For ischemic stroke, the median PPV difference between all PheValuator results and the conventional validation results was 4% (interquartile range 2-5%). Database-level PPVs in our study ranged from 0.82 (MDCD) to 0.97 (Optum EHR®), which are broadly similar to those reported from a systematic review on the validity of acute stroke diagnostic codes[11]. PPVs differential to exposure status ranged from 0.87 (THZ in CCAE) to 0.98 (ACEI in Optum EHR®) which were on average greater than the exposure agnostic PPVs (0.91 vs 0.89). In all databases in our study, sensitivity was greater in the exposure cohorts than in the overall database populations. This supports our hypothesis that patients with hypertension initiating anti-hypertensive therapy have greater healthcare utilization, and subsequently accrue more data relevant to identifying ischemic stroke than the general population. This observation provides support to the claim the PheValuator is generating accurate phenotype error estimates.

Despite its current limited use, investigators deciding whether to use PheValuator must consider the tradeoff between the value of using phenotype errors estimated using an expert-adjudicated reference standard in a small population possibly unrepresentative of their study population versus a probabilistic reference standard in their relevant study population. Our empirical example does not include analyses where phenotype errors from conventional validation were used, but it does indicate large estimate differences between estimates corrected for non-differential, database-level error and differential, exposure-level errors.